\documentclass[reprint,amsmath,amssymb,aps,pre,longbibliography,               balancelastpage,letterpaper,floatfix]{revtex4-2}
\UseRawInputEncoding
\usepackage{graphicx}
\usepackage{dcolumn}
\usepackage{bm}
\usepackage{mathtools}
\usepackage{hyperref}
\usepackage{hycolor}
\hypersetup{colorlinks=true,citecolor=[rgb]{0,0.15,0.60}, 
urlcolor=[rgb]{0,0.15,0.60}, linkcolor=[rgb]{0,0.15,0.60}, final = true}

\usepackage{enumitem}
\usepackage[rightcaption]{sidecap}
\usepackage{physics}
\usepackage{siunitx}
\usepackage{letltxmacro}
\LetLtxMacro{\OldSqrt}{\sqrt}
\newcommand{\ClosedSqrt}[1][\hphantom{3}]
{\def\DHLindex{#1}\mathpalette\DHLhksqrt}
\makeatletter
    \newcommand*\bold@name{bold}
    \def\DHLhksqrt#1#2{%
        \setbox0=\hbox{$#1\OldSqrt{#2\,}$}\dimen0=\ht0\relax%
        \advance\dimen0-0.2\ht0\relax
        \setbox2=\hbox{\vrule height\ht0 depth -\dimen0}%
        {\hbox{$#1\expandafter\OldSqrt\expandafter
        [\DHLindex]{#2\,}$}
        \lower\ifx\math@version\bold@name0.6pt\else0.4pt\fi\box2}
    }
    \renewcommand*{\sqrt}[2]
    [\ ]{\ClosedSqrt[\leftroot{-2}\uproot{1}#1]{#2}\kern0.1em} 
\makeatother

\def\ith{$i^{\mathrm{th}}$~}
\def\jth{$j^{\mathrm{th}}$~}

\def\half{{\textstyle\frac{1}{2}}}
\def\eg{{\it{e.g.,~}}} 
\def\ie{{\it{i.e.,~}}}

\def\bRb{\bar{\Rb}}
\def\ab{\mathbf{a}}
\def\bb{\mathbf{b}}
\def\eb{\mathbf{e}}
\def\fb{\mathbf{f}}
\def\Fb{\mathbf{F}}
\def\kb{\mathbf{k}}
\def\rb{\mathbf{r}}
\def\Rb{\mathbf{R}}

\def\dkb{\mathbf{dk}}
\def\bR{\bar{R}}

\def\ek{\eb_\kb}
\def\felas{\mathbf{f}^{\mathrm{elas}}}
\def\fhyd{\mathbf{f}^{\mathrm{hyd}}}
\def\felasij{\felas_{ij}}
\def\felasji{\felas_{ji}}
\def\fhydij{\fhyd_{ij}}
\def\fhydji{\fhyd_{ji}}
\def\Xij{X_{ij}}
\def\Yij{Y_{ij}}
\def\Rij{R_{ij}}
\def\Rbij{\Rb_{ij}}

\def\bRij{\bR_{ij}}
\def\bRbij{\bRb_{ij}}
\def\bRbj{\bRb_j}
\def\bRj{\bR_j}
\def\nij{\mathbf{n}_{ij}}
\def\HH{\mathbf{H}}
\def\Hk{\mathcal{H}}
\def\Hij{\HH^{ij}} 
 
\def\Hji{\HH^{ji}} 
\def\Helas{\HH_{\mathrm{elas}}}
\def\Hhyd{\HH_{\mathrm{hyd}}}
\def\Hkelas{\Hk_{\mathrm{elas}}}
\def\Hkhyd{\Hk_{\mathrm{hyd}}}
\def\Hijelas{\Helas^{ij}}
\def\Hijhyd{\Hhyd^{ij}}
\def\Hjielas{\Helas^{ji}}
\def\Hjihyd{\Hhyd^{ji}}
\def\tetij{\theta_{ij}}
\def\tetji{\theta_{ji}}
\def\tetj{\theta_{j}}
\def\sigx{\boldsymbol{\sigma}_{x}} 
\def\sigy{\boldsymbol{\sigma}_{y}}
\def\sigz{\boldsymbol{\sigma}_{z}}
\def\One{\boldsymbol{1}}
\def\tauelas{\tau_{\mathrm{elas}}}
\def\tauhyd{\tau_{\mathrm{hyd}}}
\def\Omx{\Omega_x}
\def\Omy{\Omega_y}

\def\omx{\omega_{x}}
\def\omy{\omega_{y}}
\def\omO{\omega_1}
\def\omhyd{\omega^\mathrm{hyd}}
\def\omelas{\omega^\mathrm{elas}}
\def\r{\right}
\def\l{\left}
\def\rang{\r\rangle}
\def\lang{\l\langle}
\def\AA{\mathcal{A}}
\def\earc{s} 
\def\Rey{\mathrm{Re}}
\def\kBT{k_\mathrm{B}T}

\begin{document}

\title{Exceptional Topology in Ordinary Soft Matter}

	\author{Tsvi Tlusty}
	\email{tsvitlusty@gmail.com}
	\affiliation{Center for Soft and Living Matter, Institute for Basic Science,}
	\affiliation{Physics and Chemistry Departments, Ulsan National Institute of Science and Technology, Ulsan 44919, Korea}

\date{\today}

\begin{abstract}
Hydrodynamics is shown to induce non-Hermitian topological phenomena in ordinary, passive soft matter. This is demonstrated for the first time by subjecting a 2D elastic lattice to a low-Reynolds viscous flow. The interplay of hydrodynamics and elasticity splits Dirac cones into bulk Fermi arcs, pairing exceptional points with opposite half-integer topological charges. The bulk Fermi arc is a generic hallmark of the system exhibited in all lattice and flow symmetries. Analytic model and simulations explain how the emergent singularities shape the spectral bands and give rise to a web of van Hove singularity lines in the density of states. The present findings suggest that non-Hermitian physics can be explored in a broad class of ordinary soft matter, living and artificial alike, opening avenues for topology-based technology in this regime. 
\end{abstract}

\maketitle

\section*{Introduction}
The conservation of energy in isolated Hermitian systems is a basic tenet of physics, but in practice, most systems are open, exchanging energy and information with the external world. This inherent non-Hermiticity is traditionally seen as an inevitable imperfection of realistic systems, yet recent studies revealed that it gives rise to distinctive phenomena unmatched in Hermitian physics~\cite{Bender1998}---most notably skewed spectral bands prone to symmetry breaking when exceptional points emerge~\cite{Doppler2016, ElGanainy2018, Shen2018, Moiseyev2011, Bergholtz2021}. And this discovery kicked off intensive efforts to engineer non-Hermitian systems in diverse classical and quantum settings, ranging from photonics~\cite{Zhou2018, Feng2017, Miri2019, Weidemann2020}, phononics~~\cite{Rivet2018, Zhu2018, Ma2019}, and optomechanics~\cite{Xu2016} to electronics~\cite{Schindler2011} and atomic lattices~\cite{Zhang2016}. The potential prowess of non-Hermitian technology has been demonstrated in developing new meta-materials and devices~\cite{Guo2009,Regensburger2012,Wiersig2014, Hodaei2017}. 

But an open question remains: besides sophisticated engineered systems, can one observe and utilize these so-called exotic topological phenomena in more common and natural settings?---After all, we are immersed in a dissipative, non-Hermitian world, and living systems are immanently open to exchange with the surrounding environment at all scales~\cite{Libchaber2020}. The present study offers a clear positive answer: simple model and simulations provide demonstrate for the first time non-Hermitian topological hallmarks in standard, passive elastic networks subject to ordinary viscous flow~\cite{Beatus2006,Shani2014}. Such settings are omnipresent in the overdamped low-Reynolds regime, typical to cells, macromolecules and simple soft matter systems, suggesting that non-Hermitian topology is much more common than previously realized.

Among the exotic phenomena observed in non-Hermitian materials, \textit{bulk Fermi arcs}~\cite{Zhou2018} hold a special place. In contrast to the ingrained intuition that frequency levels are closed curves, each Fermi arc is an \textit{open} isofrequency curve ending at two exceptional points. These endpoints are defects with opposite half-integer topological charges. It is important to note that bulk Fermi arcs are topological hallmarks of non-Hermiticity in the bulk spectrum of the lattice, unlike the more familiar \textit{surface} Fermi arcs induced by Weyl points in 3D Hermitian systems~\cite{Wan2011, Xu2015, Yang2018, Morali2019}. Bulk arcs have been so far elusive and were empirically observed in one photonic crystal~\cite{Zhou2018}. Recent theoretical studies predict bulk arcs in the spectra of heavy fermions~\cite{Nagai2020} and spin liquids~\cite{Yang2021}.
Strikingly, the present work finds that the elusive bulk arcs are a \textit{generic} phenomenon of driven viscoelastic matter~\cite{Beatus2012,Beatus2017,Baron2008,Beatus2007,Beatus2008} whose emergence does not require fine tuning and stems directly from the symmetry of the interactions.

An important step into the dissipative regime are recent theoretical studies~\cite{Scheibner2020, Scheibner2020a} demonstrating non-Hermitian ``odd'' elastodynamics~\cite{Shmuel2020} in a network of \textit{active} units specifically designed to exert circular forces. Likewise, non-Hermitian viscosity waves were shown to emerge in a system with ``odd viscosity"~\cite{Avron1998}. Common to these ``odd" systems are nonreciprocal elasticity or viscosity coefficients, originating from active modules that are hard to realize in microscopic settings. In sharp contrast, the present system requires no active parts, but merely standard \textit{passive} soft matter in a laminar background flow, where non-reciprocity is a direct outcome of the hydrodynamic interaction~\cite{Beatus2006,Beatus2017}. 
Thus, linking hydrodynamics and topology in soft matter can expand the realm of non-Hermitian physics to yet further fields, beyond the already rich plethora of applications developed in recent years. 

In the following Results section, we derive the dynamic equations of the hydro-elastic lattice and map its spectrum to a standard non-Hermitian form in 2D momentum space. Then, we demonstrate how the interplay of Hermitian and skew-Hermitian interactions generates bulk Fermi arcs, which are shown to be a generic hallmark of the flow-induced dynamics. Next, we discuss the topological features of these singularities and their links to the density of states. Finally, in the Discussion, we suggest possible experimental realizations and examine potential implications and future directions. 

\begin{SCfigure*}[1.0][!b]
    \centering
 \includegraphics[width=0.7\textwidth]
 {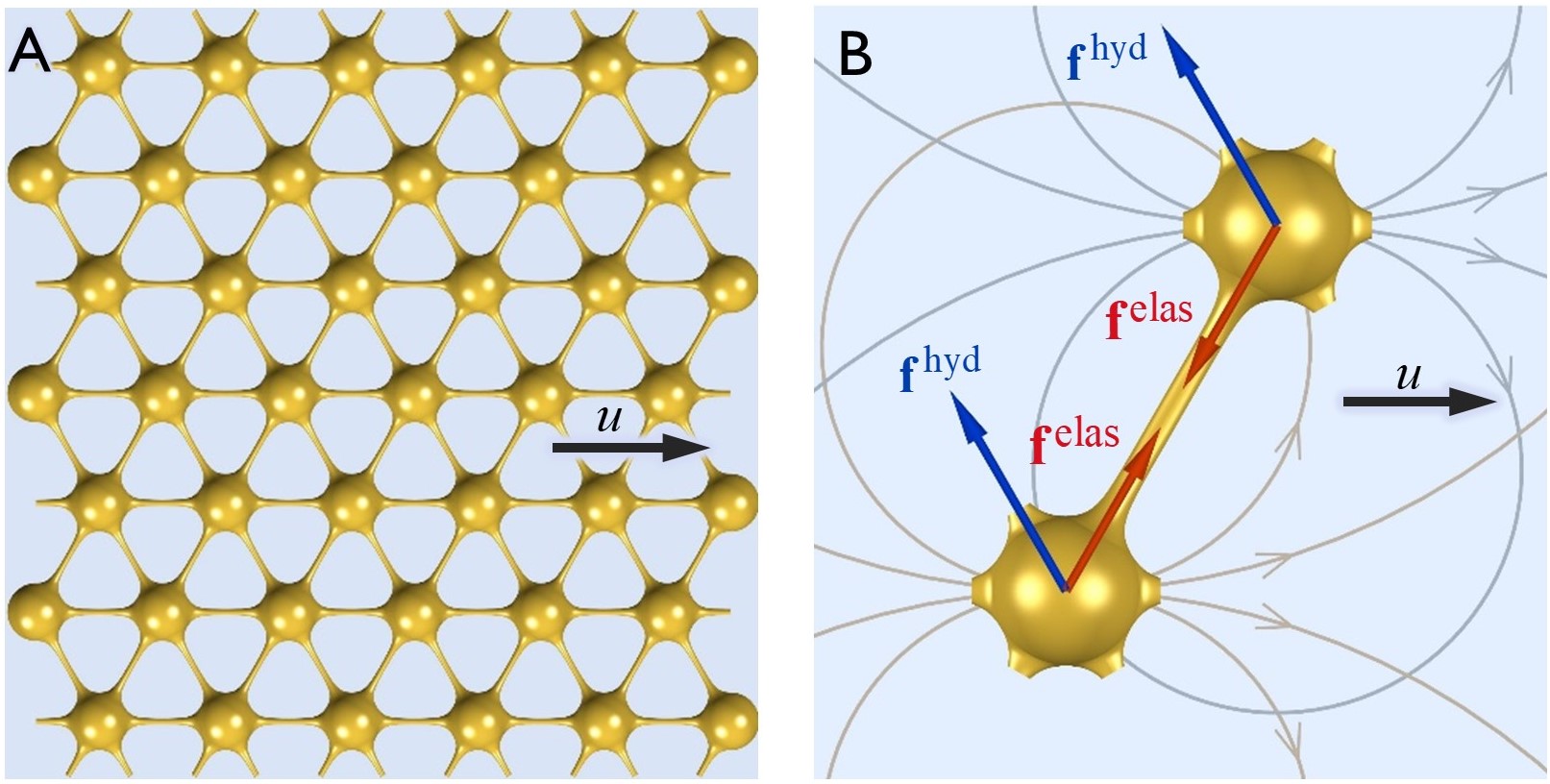}
 \caption{
(\textbf{A})  A triangular lattice made of spheres connected by elastic struts is immersed in a thin layer of viscous fluid between two walls in the $z$ direction (perpendicular to the page). The lattice is driven at a velocity $u$ relative to the fluid. \\
(\textbf{B}) A pair of moving particles induce dipolar flow fields (grey streamlines), thereby exerting on each other equal hydrodynamic forces, $\fhydij {=}~ \fhydji$ (blue arrows) in the direction of the induced flow. The elastic forces along the elastic strut are opposite, $\felasij {=} -\felasji$  (red arrows), thus conserving linear momentum.}
\label{fig:1}
\end{SCfigure*}

\section*{Results}
\noindent
\textbf{The motion of the hydro-elastic lattice.}~ 
To see how non-Hermitian topology arises in ordinary elastic matter at low-Reynolds, consider the following model system (Fig.~\ref{fig:1}A). A two-dimensional triangular lattice made of spherical particles of size $\ell$---joined by thin elastic struts of length $a$ and spring constant $\kappa$---is submerged in a viscous fluid of viscosity $\eta$ and is moving at a velocity $u$ relative to the fluid. The resulting viscous drag on each particle is $\gamma u$, where $\gamma$ is the friction coefficient. Such relative motion can be obtained in the lab by holding the lattice in a flow or by or by driving the lattice with various forces (see Discussion). 

Perturbing the surrounding fluid, the driven particles induce long-range interactions throughout the lattice~\cite{Beatus2006, Baron2008, Beatus2012, Beatus2017} (Fig.~\ref{fig:1}B). When the flow is limited to a thin sheet of viscous fluid between solid floor and ceiling, these hydrodynamics forces are dipolar ~\cite{Liron1976, Cui2004}, and the force exerted by particle $j$ on particle $i$ is 
\begin{equation}
    \fhyd_{ij} = 
    \gamma \, u \ell^2 \Rij^{-2}
    \begin{bmatrix}
    \cos{2 \tetij}\\
    \sin{2 \tetij}
    \end{bmatrix}~,
    \label{eq:fhyd}
\end{equation}
where $\Rb_i$ are the particle positions and the distance vectors are $\Rbij =\Rb_i -\Rb_j = \Rij(\cos{\tetij}, \sin{\tetij})$. The magnitude of the hydrodynamic dipole in (\ref{eq:fhyd}) scales as ${\sim} u \ell^2$. The Hookean elastic forces are proportional to the change of the length, $\Delta \Rij = \Rij - \bar{R}_{ij}$, where $\bar{R}_{ij}$ the equilibrium length of the springs ($a$ in the lattice),
\begin{equation}
    \felas_{ij} = \kappa \, \Delta\Rij
    \begin{bmatrix}
    \cos{\tetij}\\
    \sin{\tetij}
    \end{bmatrix}~.
    \label{eq:felas}
\end{equation}

Owing to the dipolar symmetry of (\ref{eq:fhyd}), the hydrodynamic forces a pair of particles exert on each other are equal, $\fhydij = \fhydji$ (because $\tetji = \pi + \tetij$). Thus, the dipolar forces break Newton's third law of momentum conservation (Fig.~\ref{fig:1}B). This is because viscous flow is an inherently open system, an \textit{effective} representation of energy and momentum transfer from hydrodynamic degrees-of-freedom to microscopic ones. To steadily move the lattice, the momentum leakage (`loss') needs to be constantly compensated by the driving force (`gain'). In contrast, the elastic forces (\ref{eq:felas}) are opposite and conserve linear momentum, $\felasij =- \felasji$. As shown below, the interplay of conservative and non-conservative forces give rise to skewed non-Hermitian topology.

Two timescales govern the dynamics of the lattice: the hydrodynamic timescale, $\tauhyd = a^3/(u \, \ell^2)$, and the elastic relaxation time, $\tauelas = \gamma/\kappa$. Their ratio is the  hydroelastic number, 
\begin{equation}
    \epsilon \equiv \frac{\tauelas^{-1}}{\tauhyd^{-1}} =
    \frac{\kappa a^3}{\gamma u \, \ell^2}~,
    \label{eq:epsilon}
\end{equation}
which controls the system's behavior: when $\epsilon \ll 1$, it is dominated by hydrodynamics, and when $\epsilon \gg 1$ by elasticity. Hereafter, we measure times in units of the hydrodynamic timescale $\tauhyd$ and frequencies in $\tauhyd^{-1}$.

The dynamical equations combine Stokes flow with Hookean elasticity, a linear regime where the correspondence between experiment and theory is well-established. In this overdamped regime, the friction force on each particle is counterbalanced by the driving force, $\Fb = F\hat{x}$, and by the hydrodynamic and elastic interactions in the lattice, 
\begin{equation}
    \gamma \dot{\Rb}_i =\Fb + 
    \sum_{j\neq i}{(\fhydij+\felasij})~,
    \label{eq:dynamic}
\end{equation}
where $\dot{\Rb}_i$ is the \ith particle's velocity (see App.~\ref{app:dynamics}). In equation (\ref{eq:dynamic}), the long-range hydrodynamic forces (\ref{eq:fhyd}) are summed over all particles, while elastic interactions (\ref{eq:felas}) are summed only among neighbors connected by struts. Owing to the lattice parity symmetry, the sums of interactions vanish at steady-state, when the lattice traverses uniformly at a velocity $u= F/\gamma$. \\

\begin{figure*}[!htb]
\centering
\includegraphics[width=1.0\textwidth]
{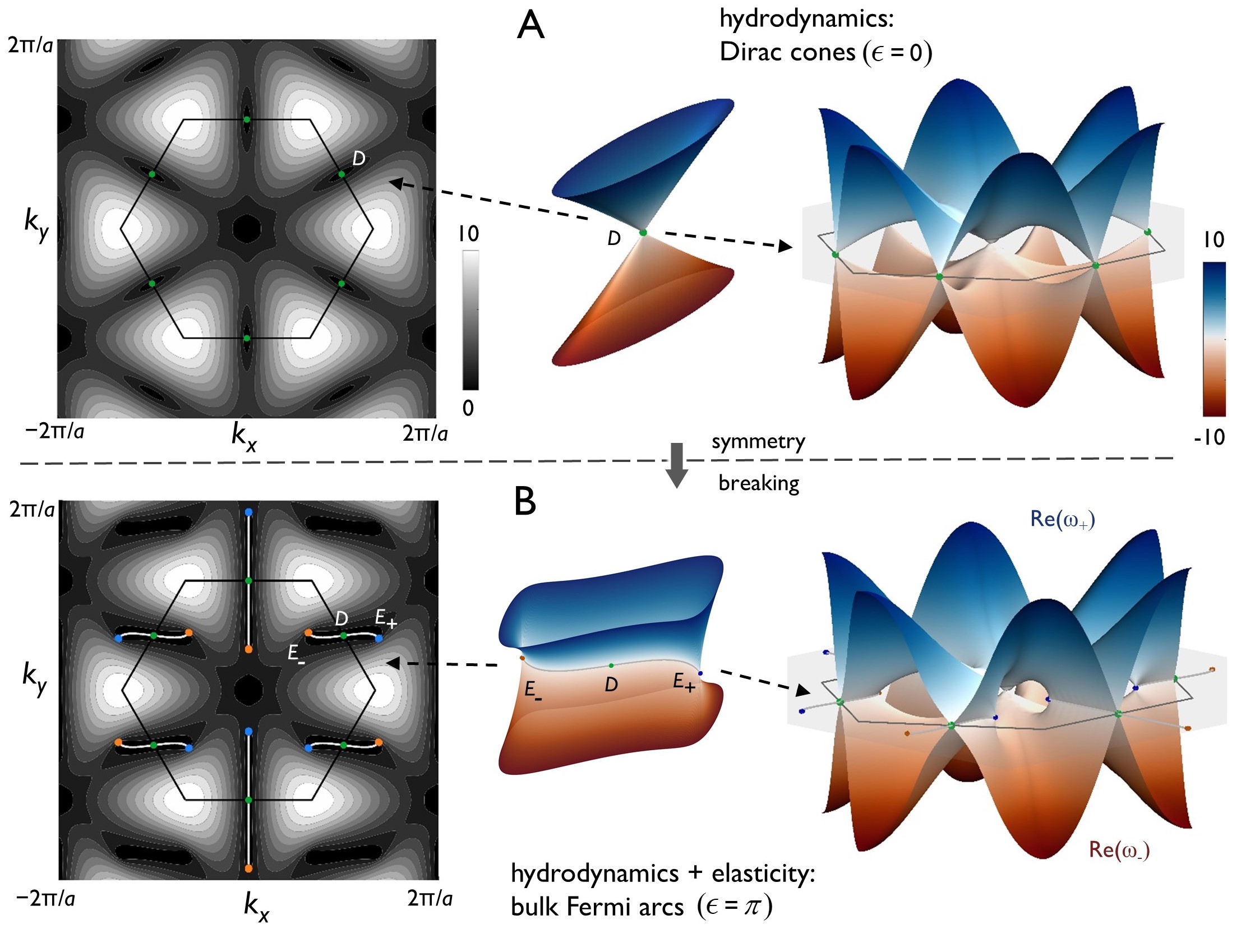}
\caption{
(\textbf{A}) A purely hydrodynamic system, $\epsilon = 0$. 
Left: The operator $\Hk =\Hkhyd$ is Hermitian with real frequency bands $\omega_+ = -\omega_- = \sqrt{\Omx^2+\Omy^2}$. The spectrum exhibits six Dirac points (green, one denoted as D) on the boundary of the Brillouin zone (black hexagon). 
Middle: At each Dirac point, the bands merge, forming a graphene-like double-cone, ``diabolo" shape. 
Right: the 3D shape of the frequency bands in the first Brillouin zone (grey hexagon), showing the double-cones, which are halved by the zone's boundary.
(\textbf{B}) At $\epsilon \neq 0$, the symmetry is broken when $\Hk$ includes a skew-Hermitian component, $\epsilon\Hkelas$. 
Left: The real part of the bands, $\Re(\omega_+) = - \Re(\omega_-)$, drawn for  $\epsilon = \pi$, exhibits six bulk Fermi arcs (white lines) in the first Brillouin zone (the whole spectrum is shown in Fig.~\ref{fig:allspec-S}). Each arc splits from a Dirac point (green) and joins two Exceptional Points (ExPs) with topological charges $\pm \half$ (orange and light blue, $E_+$ and $E_-$ are two ExPs that split from $D$). 
Middle: the real parts of the bands merge along the Fermi arc, forming a double-wedge shape. 
Right: 3D shape of the the frequency bands in the first Brillouin zone (grey hexagon), showing the double-wedges (halved by the zone's boundary).}
\label{fig:2}
\end{figure*}

\noindent \textbf{Dynamics in momentum space.}~
The coaction of elastic and hydrodynamic forces excites collective motion in the lattice. Expanding the dynamics (\ref{eq:dynamic}) in small deviations $\rb_j$ of the particles from their steady-state positions in the moving lattice $\bRbj$, we find that the collective modes are plane waves $\rb_j = \ek \exp [i(\kb \cdot \bRbj- \omega t)]$ (App.~\ref{app:linear_expansion}-\ref{app:momentum_space}). The 2D polarization of the wave, $\ek$, is an eigenstate of a Schr\"odinger-like equation with an eigenfrequency $\omega$,
\begin{equation}
        \Hk \, \ek = \omega \, \ek~, 
        \label{eq:schrodinger}
\end{equation}
where the operator $\Hk$ is a momentum-space representation of the forces in the lattice. The ``Hamiltonian" $\Hk$ is a $2 {\times} 2$-matrix, which sums the hydrodynamic and elastic interactions with a relative weight $\epsilon$. In the basis of left- and right-circular polarizations, we obtain
\begin{align}
   & \Hk  =  ~\Hkhyd + \epsilon\, \Hkelas~,~  
    \nonumber\\
     \text{with} \qquad 
     &  \Hkhyd   =  ~\Omx \sigx + \Omy \sigy~, 
    \label{eq:Hk}\\
    \text{and} \qquad  &\Hkelas  =  -i \l(\omx \sigx + \omy \sigy + \omO \One \r)~,
    \nonumber
\end{align}
where $\sigx$, $\sigy$ and $\One$ are Pauli's and the unity matrices. The frequencies $\Omx$, $\Omy$, $\omx$, $\omy$ and $\omO$, are the Fourier sums of the interactions---all real by the parity symmetry of the lattice (see App.~\ref{app:momentum_space}). The hydroelastic operator $\Hk$ (\ref{eq:Hk}) is analogous to the Hamiltonian of spin-$\half$ particles in a complex magnetic field with damping~\cite{Landau1935}, and this spinor-like nature shows in the spectrum, as discussed below. Equation~(\ref{eq:schrodinger}) with its non-Hermitian Hamiltonian $\Hk$~(\ref{eq:Hk}) is standard and appears in numerous 2D topological systems. However, the symmetry of $\Hk$ in momentum space is non-trivial and leads to the formation of bulk Fermi arcs. \\

\noindent \textbf{Symmetry: Parity and Hermiticity.}~
The hydrodynamic operator $\Hkhyd$ in (\ref{eq:Hk}) is Hermitian, a sum of products of Hermitian Pauli matrices and real numbers. Likewise, the elastic operator $\Hkelas$ is skew-Hermitian (\ie a Hermitian operator multiplied by $i$), 
\begin{equation*}
    \Hkhyd = \Hkhyd^\dag ~,~\Hkelas = - \Hkelas^\dag~.
\end{equation*}
Note that the hydrodynamic forces do not conserve momentum, but the effective hydrodynamic operator is Hermitian. This is because, in the low-Reynolds regime, the sum of the forces is proportional to the velocity (the drag force in (\ref{eq:dynamic})). Mathematically speaking, the imaginary unit factors of the time derivative ($i \omega$) and the spatial derivative ($i \kb $) cancel each other. For the same reason, the effective elastic operator is skew-Hermitian, reflecting the overdamped nature of elastic modes in this regime. As for parity symmetry in $\kb$-space, the hydrodynamic part is odd and the elastic part is even, 
\begin{equation*}
    \Hkhyd(-\kb) =  - \Hkhyd(\kb)~,~
 \Hkelas(-\kb) =  + \Hkelas(\kb)~.
\end{equation*}
This follows from the parity of the interactions (\ref{eq:fhyd},\ref{eq:felas}): $\Omx$ and $\Omy$ are odd functions of $\kb$ whereas $\omO$, $\omx$ and $\omy$ are even (see App.~\ref{app:momentum_space}). It is important to note that unlike ``odd" systems \cite{Scheibner2020,Avron1998}, $\Hk$ does not introduce any nonreciprocal coefficients of elasticity or viscosity. Here, non-Hermiticity is simply the outcome of hydrodynamics. \\

\noindent\textbf{The spectrum: Dirac cones and Fermi arcs.}~
The interplay of odd, Hermitian hydrodynamics and even, skew-Hermitian elasticity brings about distinctive topological signatures (Fig.~\ref{fig:2}). The spectrum of equations (\ref{eq:schrodinger},\ref{eq:Hk}) exhibits two eigenfrequency bands,
\begin{equation}
\omega_{\pm} = 
- i \epsilon \,\omO \pm \sqrt{\nu_+ \nu_-}~,
\label{eq:specturm}
\end{equation}
where $\nu_{\pm} = (\Omx - i \epsilon \omx) \pm i ( \Omy - i \epsilon \omy )$, and all frequencies are measured in units of $\tauhyd^{-1}$. The corresponding polarization eigenstates are (App.~\ref{app:spectra})
\begin{equation}
      \ek^{\pm} =
      \frac{1}
      { \sqrt{ \l| \nu_+ \r| + \l| \nu_- \r|} }
      \begin{bmatrix}
      \sqrt{\nu_{\pm}}\\
      \sqrt{\nu_{\mp}}
      \end{bmatrix}~.
      \label{eq:eigenvectors}
\end{equation}

Without elastic forces (Fig.~\ref{fig:2}A), a purely hydrodynamic system ($\epsilon = 0$) exhibits real spectrum of propagating phonon-like waves~\cite{Beatus2006, Baron2008}, $\omega_+ = -\omega_- = (\Omx^2+\Omy^2)^{1/2}$. On the edge of the Brillouin zone there are six Dirac points where the hydrodynamic interaction vanishes, $\Omx = \Omy = 0$. At a Dirac point, negative and positive bands kiss, $\omega_+ = \omega_- = 0$, forming a graphene-like double cone~\cite{Yarkony1996} (App.~\ref{app:Dirac_points}). 

\begin{figure}[t!]
    \centering
    \includegraphics[width=1.0\columnwidth]    {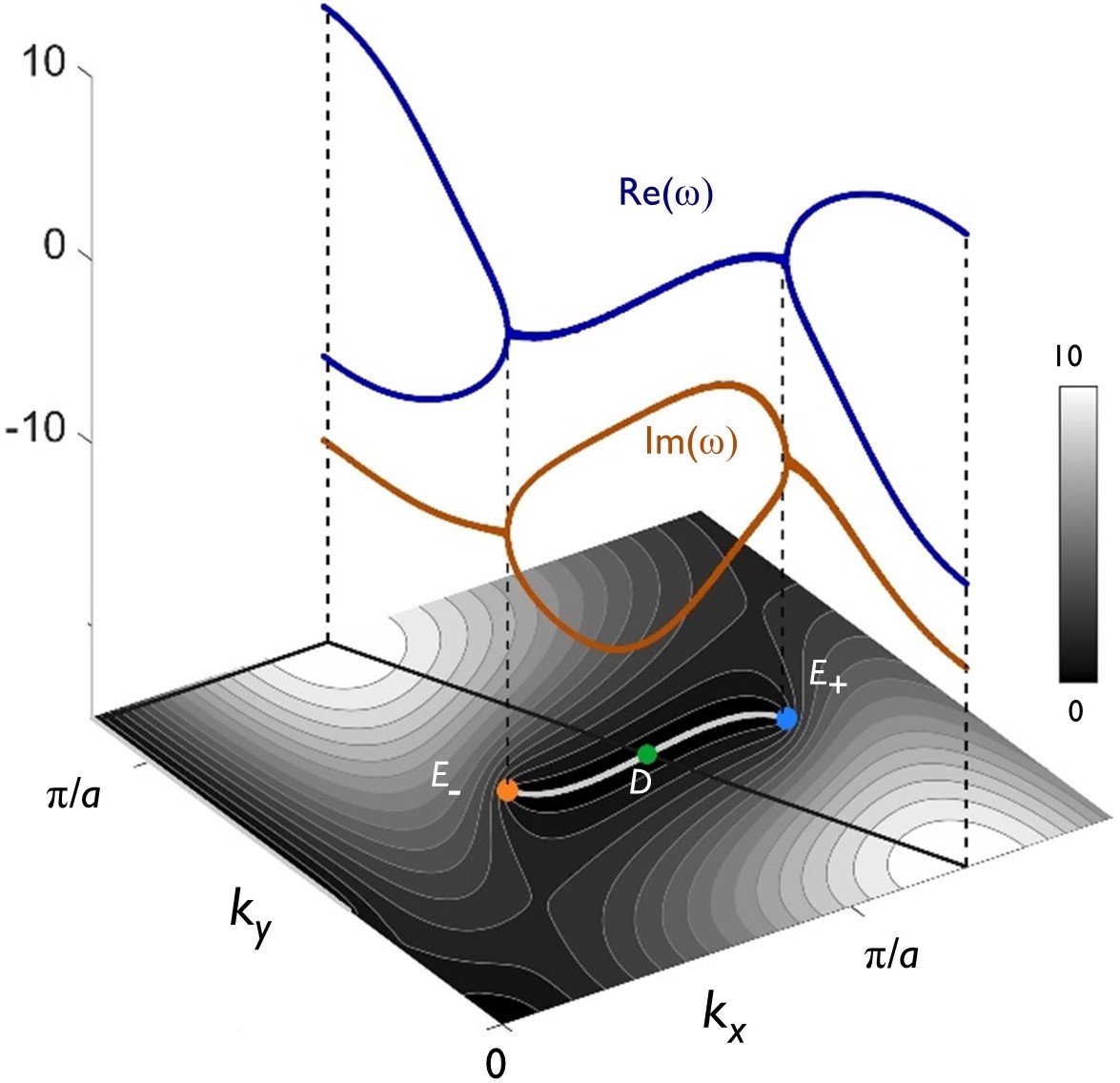}
    \caption{The double branching transition in the real (blue) and imaginary (red) parts of the bands at the exceptional points (\ref{eq:branching}). Zoom on the upper right bulk Fermi arc in Fig.~\ref{fig:2}B.}
    \label{fig:3}
\end{figure}

The introduction of skew-Hermitian elasticity at non-zero $\epsilon$ breaks the symmetry. Fig.~\ref{fig:2}B shows the real part of the bands, $\Re(\omega_+) = - \Re(\omega_-)$, for a hydroelastic number $\epsilon = \pi$ (the whole spectrum is shown in Fig.~\ref{fig:allspec-S}). 
Six bulk Fermi arcs---four S-shaped and two vertical lines---emerge from the Dirac points. Along the arcs, the real parts of the bands merge at $\Re(\omega_+) = \Re(\omega_-) = 0$. Each arc is therefore an \textit{open-ended} isofrequency contour joining a pair of \textit{isolated} exceptional points (ExPs). As shown below, these points are topological defects of opposite $\pm \half$ charges (App.~\ref{app:EPs_and_FermiArcs}).\\

\noindent\textbf{Singularities and bifurcations.}~
At the ExPs, the spectral bands (\ref{eq:specturm}) and their corresponding eigenstates (\ref{eq:eigenvectors}) simultaneously coalesce (Fig.~\ref{fig:2}B): the bands are purely imaginary, $\omega_+ = \omega_- = - i \, \epsilon \, \omO$, and the coalescing eigenstates, $\ek^{+} = \ek^{-}$, are either right- or left-circular polarizations, reflecting the chirality of the topological charges (see App.~\ref{app:EPs_and_FermiArcs}). ExPs occur where the determinant in (\ref{eq:specturm}) vanishes, ($\nu_+ = 0$ or $\nu_- = 0$) at
\begin{equation}
    \frac{\Omy}{\omx} = -\frac{\Omx}{\omy} 
    = \pm \epsilon ~.
    \label{eq:exceptional}
\end{equation}
At these double branching singularities, of both $\Re(\omega)$ and $\Im(\omega)$, the spectrum becomes gapless. In contrast to the Dirac points whose eigenspaces are two-dimensional, the eigenstates at the ExPs are parallel, signifying a reduction of the eigenspace dimension to one. 

Tuning the hydroelastic number $\epsilon$ advances the ExPs along 1D trajectories, from the Dirac points at $\epsilon = 0$ to the corners or the center of the Brillouin zone at $\epsilon = \infty$ (App.~\ref{app:EPs_and_FermiArcs}). Along these trajectories, the bands exhibit square-root singularities (Fig.~\ref{fig:3}), 
\begin{equation}
\omega_{\pm} = - i \epsilon \,\omO \pm 
\sqrt{\l(\omx^2 +\omy^2 \r)\l( \earc^2 - \epsilon^2 \r)}~, 
\label{eq:branching}
\end{equation}
where $\earc \equiv \Omx /\omy = -\Omx/\omy$ is a 1D coordinate defined by (\ref{eq:exceptional}). The Fermi arc is the branch-cut of the square root (\ref{eq:branching}) stretching between the bifurcation transitions at the ExPs, $\earc = \pm \epsilon$. The square-root singularity reflects strong level repulsion between the bands, compared to the linear opening of the gap at the Dirac cone~\cite{Miri2019}. \\

\begin{figure}[b!]
\centering
\includegraphics[width=1.0\columnwidth]
{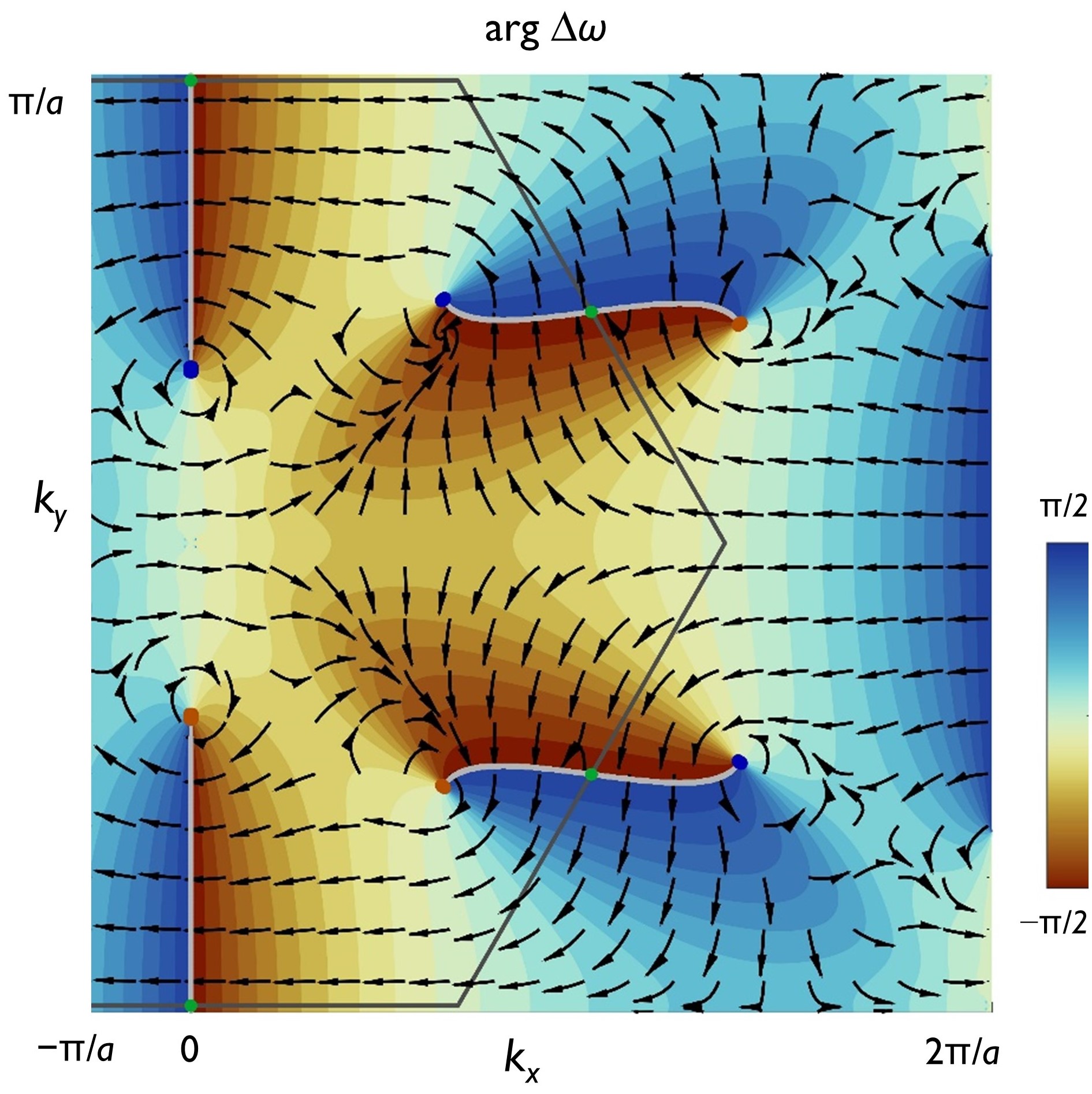}
\caption{The vorticity of the band-gap, 
$\Delta \omega = \omega_+ - \omega_-$ (\ref{eq:vorticity}), for $\epsilon = \pi$. The argument of the band-gap, $\arg{\Delta\omega}$, is color-coded and arrows denote its gradient field, $\nabla_\kb (\arg{\Delta\omega})$. Hexagonal black line shows the Brillouin zone boundary, with Dirac points (green) and exceptional points with $\pm \half$ charges (orange and blue). }
\label{fig:4}
\end{figure}

\begin{figure*}[htb!]
\centering
\includegraphics[width=1.0\textwidth]{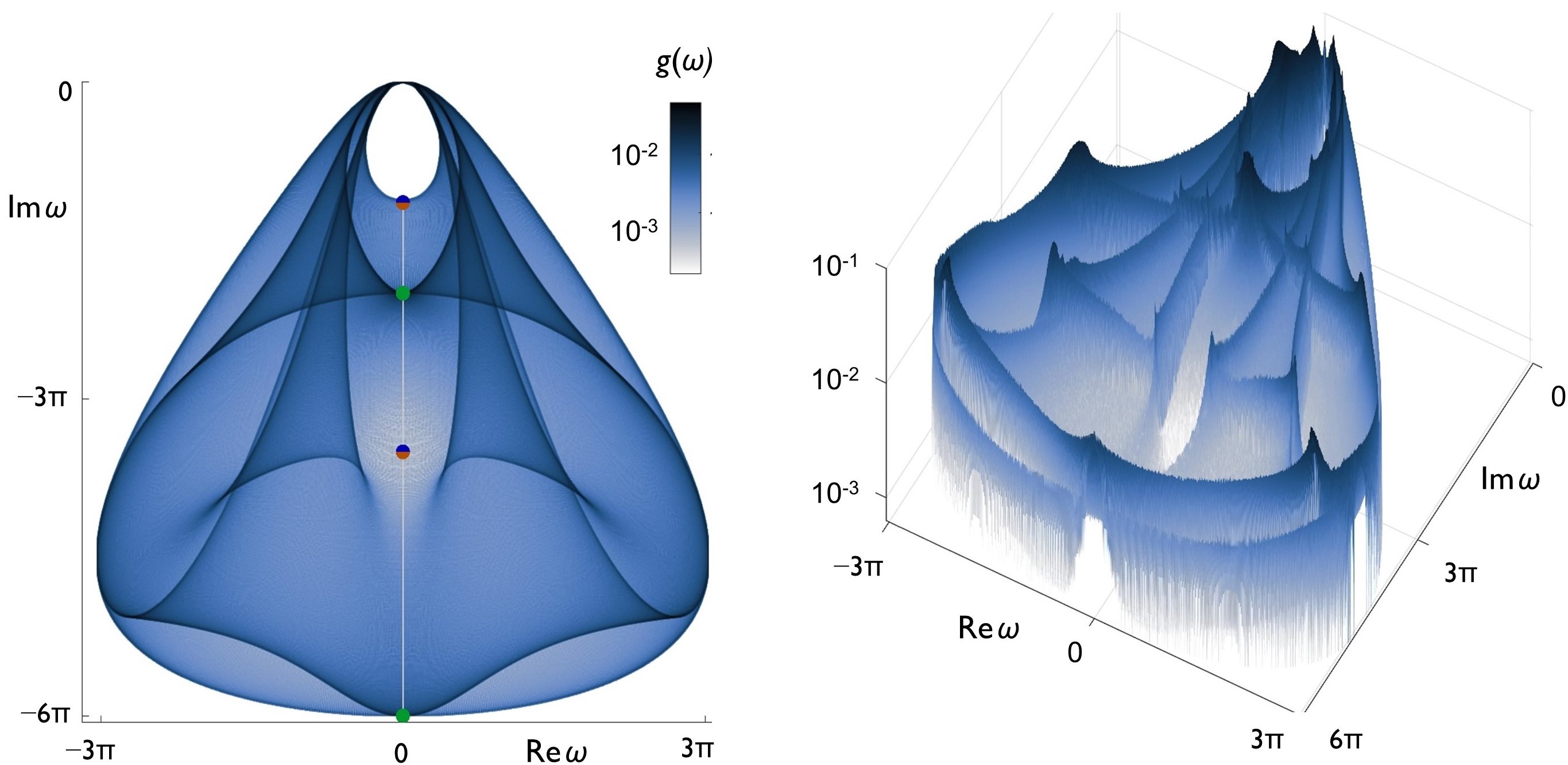}
\caption{
Left: The density of states $g(\omega)$ plotted in log-scale, showing the Dirac points (green circles) and the exceptional points (blue/orange circles), all connected by bulk Fermi arcs (white line) at $\Re \omega = 0$.  
Right: 3D representation of $g(\omega)$.}
\label{fig:5}
\end{figure*}

\noindent\textbf{Bulk Fermi Arcs are generic.}~
As mentioned, the \textit{bulk} Fermi arcs are topological signature of non-Hermiticity~\cite{Zhou2018} in the driven elastic lattice---unrelated to the more common \textit{surface} Fermi arcs induced by Weyl points in 3D Hermitian systems \cite{Wan2011, Xu2015, Yang2018, Morali2019}. The bulk arcs are a direct outcome of the broken symmetry, at $\epsilon \neq 0$, when hydrodynamics and elasticity with opposite symmetries are mixed. Thus, the bulk arcs are a \textit{generic} feature of the system, which are indeed observed in other lattice symmetries (Fig.~\ref{fig:allS-square}) and for all directions of the driving flow. This is formally shown by a small-$\epsilon$ expansion around the Dirac cone of a purely hydrodynamic system (App.~\ref{app:generic}), demonstrating that solutions to the ExP double condition (\ref{eq:exceptional}) always exist. Thus, the pair of ExPs and the bulk arc that stretches between them are generic.\\

\noindent\textbf{Topological charges.}~
The ExPs are topological defects: a closed eigenfrequency loop encircling an ExP cannot shrink to a point without passing through the ExP. The corresponding topological charge can be found from the vorticity of the band-gap~\cite{Shen2018}, $\Delta \omega = \omega_+ - \omega_-$ (Fig \ref{fig:4}), 
\begin{equation}
     \mathcal{V} = -\frac{1}{2\pi} \oint
     \dkb \cdot \nabla_\kb \l( \arg{\Delta \omega} \r)~.
     \label{eq:vorticity}
\end{equation}
 Since the band-gap is $\Delta \omega = 2 \sqrt{\nu_+ \nu_-}$  (\ref{eq:specturm}), the vorticity is
 \begin{equation}
     \mathcal{V} = -\frac{1}{4\pi} \oint 
     \dkb \cdot \nabla_\kb \l( \arg{\nu_+} + \arg{\nu_-} \r)
     = \pm \frac{1}{2}~,
     \label{eq:vorticityCirc}
 \end{equation}
with a sign corresponding to the left- or right-handed chiralities of the gradient field (App.~\ref{app:vorticity}). The vorticity is determined by the branch-cuts, where the jumps of $\arg\nu_\pm$ by $\pm 2 \pi$ yield the topological charges $q_{\pm} = \pm \half$. The charges and their opposite chiralities originate from the square-root singularity of the Riemann surface  (\ref{eq:branching}), and reflect the spinor-like nature of the polarization eigenstates, which accumulate a $\pm \pi$ phase when circling the ExP and passing through the branch-cut of the arcs. \\

Likewise, the charges can be computed from the integral of the Berry phase~\cite{Berry1984} (see App.~\ref{app:Berry}). Berry's connections are the vectors
\begin{equation*}
    \AA_\pm(\kb) = i \,\ek^{\pm\ast}
    \nabla_\kb \ek^{\pm}~,
\end{equation*}
and Berry's phases $\gamma_\pm$ are the loop integrals
\begin{equation}
    \gamma_\pm = \oint \AA_{\pm}(\kb) \cdot \dkb = \pm \pi~.
    \label{eq:BerryPhase}
\end{equation}
The corresponding charges, 
$q_\pm = \gamma_{\pm}/(2 \pi) = \pm \half$,
are determined by the jump of the phase (\ref{eq:BerryPhase}) at the branch-cut, as in the case of the vorticity (\ref{eq:vorticity}).\\

\noindent\textbf{The density of states and its singularities}.~
Projecting the Riemann surfaces of the spectral bands $\omega_\pm$ onto the complex frequency-plane reveals the density of states $g(\omega)$ (Fig.~\ref{fig:5}), 
\begin{equation}
    g(\omega) = \l( \frac{a}{2 \pi} \r)^2
    \int{ d^{2}\kb \, \delta \l(\omega - \omega(\kb)\r)}~.
\end{equation}
The Dirac points (green) split between the two banks of the branch cut. Notable are sharp-edged ridges of the density merging at logarithmically diverging summits, akin to van Hove singularities in Hermitian systems~\cite{VanHove1953}. Level repulsion shows in low density of states around the ExP singularities in Fig.~\ref{fig:5}. The analytic density of states $g(\omega)$ is similar to the one obtained from spectra of simulated lattices, with deviations owing to finite size and boundary conditions (Fig.~\ref{fig:simulations-S}). \\

\section*{Discussion}

\noindent\textbf{Realizations and physical limitations.~}
Four physical conditions determine the experimentally accessible regime:
\begin{enumerate}[labelwidth=18pt, leftmargin = 0pt, itemindent = 22pt, label=\Roman*.~]
\item  The topology is most notable when the elastic and hydrodynamic forces are comparable, around $\epsilon \sim 1$. Using (\ref{eq:epsilon}), this condition can be written as $\kappa \sim  \eta u (\ell/a)^3$ (with Stokes law $\gamma \sim \eta \ell$).
\item  The Reynolds number $\Rey$ must be kept in the overdamped non-inertial regime, $\Rey = \rho u \ell/\eta \le \Rey_\ast$, with an upper bound $\Rey_\ast \simeq \numrange{e-2}{e-1}$. Using condition I, this amounts to $\kappa \le \Rey_\ast[\eta^2/(\rho \ell)](\ell/a)^3$.
\item To avoid lattice ``melting", the thermal fluctuations should be kept small. A Lindemann-like criterion implies that the fluctuations are smaller than the particle size, $\lang \rb_i^2 \rang \sim (\kBT/\kappa) \ln(L/a) \le \ell^2$, where $\kBT$ is the thermal energy scale and the fluctuations of the 2D elastic lattice increase logarithmically with size $L$~\cite{Safran2018}. This condition is recast as a lower bound, $\kappa \ge (\kBT/\ell^2) \ln(L/a)$.
\item The dynamics should be fast enough to allow adequate collection of statistics, $\eta\ell/\kappa \sim \tauelas \sim \tauhyd \le \tau_\ast$, with an upper bound taken as $\tau_\ast \sim \SI{10}{\s}$~\cite{Beatus2006, Beatus2017}. This sets another lower bound on the elasticity, $\kappa \ge \eta\ell/\tau_\ast$.
\end{enumerate}

\begin{figure}[b!]
\centering
\includegraphics[width=1.0\columnwidth]{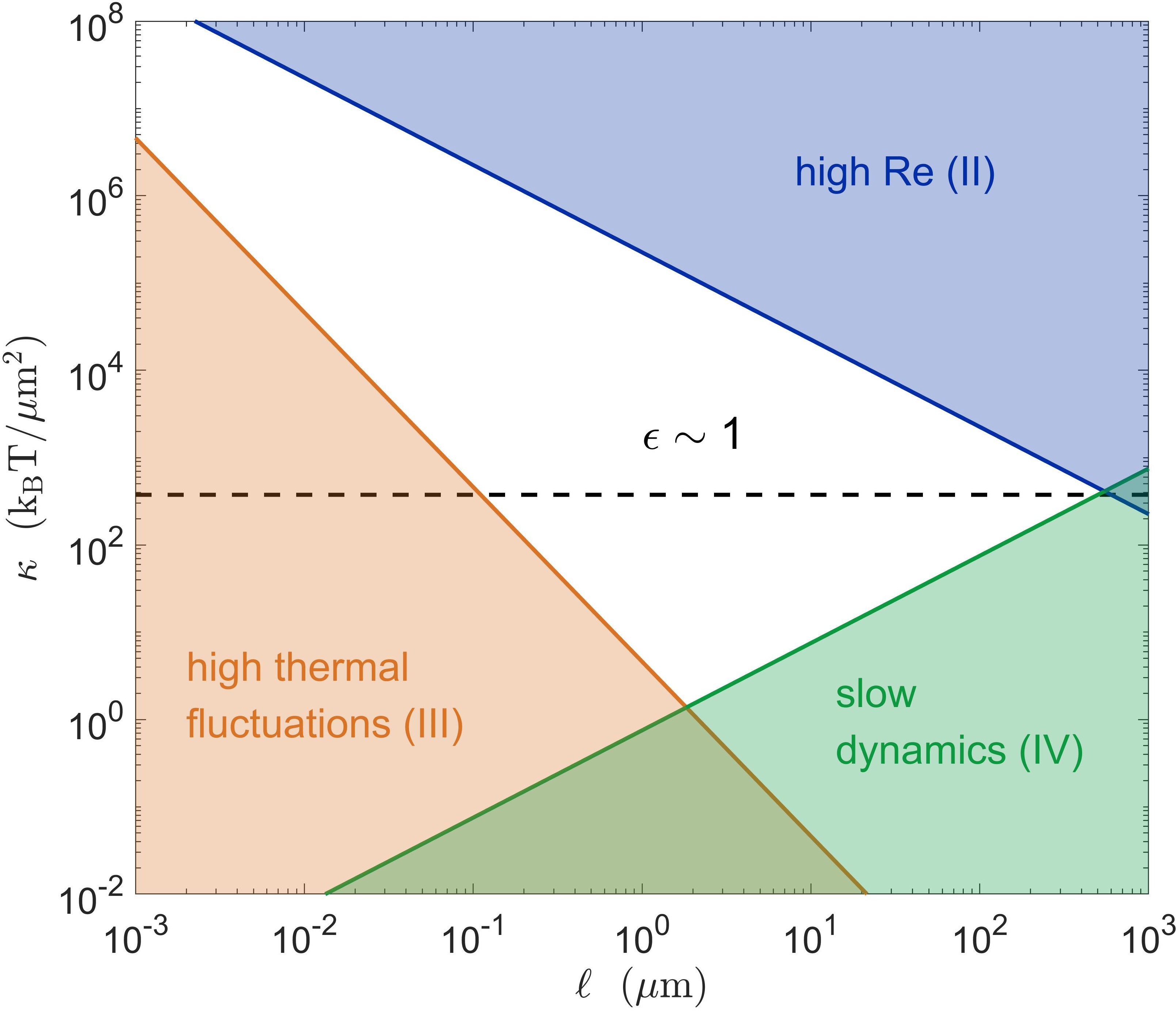}
\caption{
The physical limitations I-IV in the $\ell{-}\kappa$ plane, drawn for for typical values of a microfluidic system~\cite{Beatus2017}: velocity $u = \SI{500}{\um\per\s}$, viscosity $\eta = \SI{30}{cP}$, geometric factor $(\ell/a)^3 = 0.1$, and lattice size $(L/a)^2 = \num{e4}$. The regions excluded by bounds II-VI are shaded, and condition I ($\epsilon \sim 1$) is the dashed line. The accessible region is in the remaining white domain.} 
\label{fig:6}
\end{figure}
Examination of the bounds in the $\ell{-}\kappa$ plane (Fig.~\ref{fig:6}) shows that the regime they define is easily accessible in standard soft matter settings. The limits are drawn for the parameters of a prototypical microfluidic system~\cite{Beatus2017}. We see that all conditions I-IV are met by the optimal spring constant $\kappa \sim \SIrange{e2}{e3}{\kBT\per\um^2}$ (from condition I), corresponding to interaction energies of order $\SIrange{e2}{e4}{\kBT}$ between micron-sized particles.

Physical settings of microscopic bodies whose interactions are of order ${\sim}\SI{e3}{\kBT}$ are characteristic to various soft matter systems, in particular colloids \cite{Safran2018}. For example, spring constants of about ${\sim}\SIrange{e2}{e3}{\kBT\per\um^2}$ were measured in colloidal gels~\cite{Dinsmore2006}, small lattices~\cite{Park2014} and 2D colloidal crystals \cite{Buttinoni2021,Zhang2004,Chen2011,Mao2013}. Note that elasticity can be replaced by any other conservative central force, such as magnetic~\cite{Keim2004} and electrostatic~\cite{Kelleher2017} interactions---all retain the Hermiticity and parity symmetries of elasticity and enter the equations of motion as an effective spring constant. Other relevant systems include lattices of DNA-coated colloids~\cite{Wang2015, Rogers2016}, colloidal lattices embedded in gels, and self-assembling microfluidic crystals~\cite{Lee2010, Uspal2014}.

To observe the exceptional topology, such soft matter crystals should be subjected to viscous flow, for example in a quasi-2D microfluidic channel. The lattices can be driven by the surrounding flow, slowed down by friction~\cite{Beatus2017} or held in place by boundary conditions or external forces, for example, by dynamic holographic optical tweezers~\cite{Curtis2002}. What matters is the velocity of the lattice relative to the embedding fluid, which induces the hydrodynamic interactions. The geometry of the lattice can be easily adapted. In our model system (Fig.~\ref{fig:1}A), we assumed for simplicity that the particles are joined by thin struts. However, the struts can be replaced by other particles, forming for example kagome, honeycomb or square-kagome lattices. This would augment the non-Hermitian spectrum with additional optical branches. Moreover, long-range magnetic or electrostatic interactions obviate the need for direct mechanical connection among the particles. So one can leave out the struts and form, for instance, large 2D lattices of magnetic colloids~\cite{Zahn2000}, disks~\cite{Grzybowski2002}, hydrogel particles~\cite{Hwang2008}, or droplets~\cite{Katsikis2015},  with effective springs induced by magnetic dipole-dipole interactions \cite{Keim2004}. The steady state configuration of such a lattice is governed by the confining boundaries, which can be crafted to allow the passage of viscous flow. 
\\

\noindent\textbf{Outlook and summary.}~
The proposed non-Hermitian model system can also be realized also in the  sub-micron regime. Shrinking the system's dimensions (Fig.~\ref{fig:6}) 
tightens the thermal stability bound on $\kappa$ (condition III), which approaches the optimal value (condition I) at a particle size of about $\ell \sim \SI{100}{\nm}$. But the bounds may be relaxed by increasing $\eta u$ (shifting up the dashed line), for example using higher pressures in hardened microfluidic chambers~\cite{Beatus2012, Shani2014}. Tentative candidate systems in this regime include lattices made of DNA~\cite{Thaner2015, Rogers2016}, protein~\cite{Sinclair2011, BenSasson2021, Zhang2020} and patchy nano-particles~\cite{Zhang2004, Liljestroem2015, Li2020}. At the other extreme, the system can be expanded up to the sub-millimetric regime, as long as the effective springs remain soft. Minding the omnipresence of both periodic lattices \cite{Messner2010, Charrier2019} and low-Reynolds flows \cite{Purcell1977, Brody1996, Lauga2009} in biology and biotechnology, one may speculate that non-Hermitian topological phenomena are also widespread in these realms. 

To sum, the reported skewed topology with bulk Fermi arcs ending at isolated exceptional points was recently observed in a photonic crystal with radiation loss~\cite{Zhou2018}. Here, we proposed a blueprint a controllable realization of this exotic topology in ordinary, passive soft matter, where the analogue of radiation loss is viscoelastic dissipation.  The performed analyses showed the ubiquity of exceptional points and bulk Fermi arcs in generic elastic lattices driven by viscous flow. This suggests that non-Hermitian topology---which is robust against defects and noise---can be used to develop devices in the low-Reynolds regime. While these directions are intriguing, our main objective here was to demonstrate that simple viscoelastic systems can form an easily-accessible playground to investigate fundamental features of topological matter in the overdamped low-Reynolds regime, typical to soft and living matter. Future directions include examination of skin modes~\cite{Bergholtz2021} and 3D geometries, where the long-range hydrodynamic interaction is much stronger than in 2D. In addition, one could explore composite architectures combining passive and active soft matter, also at the nanoscale \cite{Jee2018,Jee2018a}.  

\begin{acknowledgments}
This work was supported by the Institute for Basic Science, Project Code IBS-R020. 
I thank Tsevi Beatus, Roy Bar-Ziv, Sam Safran, Hyuk Kyu Pak, Vincenzo Vitelli, William Irvine and their coworkers for essential comments. 
\end{acknowledgments}

\appendix

\section{Derivation of the model}

\subsection{Dynamics of hydroelastic lattices}
\label{app:dynamics}
The following is a brief description of the derivation. An elastic lattice is moving in the $x$-$y$ plan of a thin 2D fluid layer at a velocity $u$ relative to the fluid. In this quasi-2D geometry, the narrow dimension is $z$ (perpendicular to the page in Fig.~\ref{fig:1}).  The lattice is made of particles of size $\ell$ joined by thin elastic rods of average length $a$ and spring constant $\kappa$. The viscous drag on each particle is $\gamma u$, where $\gamma$ is the friction coefficient (inverse of the mobility). Therefore, this non-equilibrium steady-state of uniformly moving lattice requires driving forces $\Fb$ (force per particle) that inject momentum and energy to compensate for the dissipative friction forces.

The particles' motion with respect to the fluid induces dipolar perturbations with a velocity field decaying as the inverse square of the distance, ${\sim} u \, (\ell/r)^2$, and this dipolar flow fields give rise to collective hydrodynamic interactions \cite{Beatus2006, Baron2008, Beatus2012, Beatus2017}.  The hydrodynamic force $\fhydij$ exerted by the \jth particle on the \ith particle is (in $x,y$ components),
\begin{align}
    \fhyd_{ij,\, x} & = \gamma \Lambda \sum_{j \neq i}
    {\frac{\l(\Xij^2 - \Yij^2\r)}
    {\l(\Xij^2+\Yij^2\r)^2}}  = &
    \gamma \Lambda \sum_{j \neq i}
    {\frac{\cos{2 \tetij}}{\Rij^2}}~,
        \label{eq:fhyd-M}\\
    \fhyd_{ij,\,y} & = \gamma \Lambda \sum_{j \neq i}
    {\frac{2\Xij \Yij}{\l(\Xij^2+\Yij^2\r)^2}}  = &
    \gamma \Lambda \sum_{j \neq i}
    {\frac{\sin{2 \tetij}}{\Rij^2}}~, \nonumber
\end{align}
where the positions of the dipoles are $\Rb_i=(X_i,Y_i)$, and $\Rbij =\Rb_i -\Rb_j = (\Xij,\Yij)$ are the distance vectors. In polar coordinates, $\Rbij = (\Rij,\tetij)$, where $\Rij = |\Rbij|$ and $\tetij$ the angle. Equation (\ref{eq:fhyd}) is a compact form of (\ref{eq:fhyd-M}).

The coupling constant $\Lambda$ scales as the strength of the dipoles, $\Lambda \sim  u \, \ell^2$, where $\ell$ is the size of the particle, and $u$ its velocity \textit{relative to the fluid}. This defines the typical timescale $\tauhyd$ of the hydrodynamic interaction,
\begin{equation}
    \tauhyd \equiv  \frac{a^3}{\Lambda}
    =\frac{a^3}{u \, \ell^2} ~,
\end{equation} 
the time it takes a perturbation to traverse a distance $a$ at a sound velocity $c_s \sim \Lambda/a^2$ (in the continuum long wavelength limit). The timescale $\tauhyd$ depends on the physical forces driving the hydrodynamic interaction. For example, in a lattice of particles sedimenting in a quasi-2D fluid, the relative velocity scales as $ u \sim \Delta\rho\, g \ell^2/\eta$, where $\Delta\rho$ is the density difference and $\eta$ the fluid's viscosity. In the quasi-2D flow of squeezed droplets, the coupling is $\Lambda = \ell^2 K u$, where $K = u_{\rm particle}/u_{\rm fluid} \le 1$ is the ratio of the velocities of particle and the surrounding fluid that drags it in the channel~\cite{Beatus2006,Beatus2017}.

The elastic Hookean forces are described by harmonic springs with constant $k$ and equilibrium length $\bar{R}_{ij}$, the lattice constant $a$. The elastic forces are proportional to the change of the length, $\Delta \Rij = \Rij - \bar{R}_{ij}$,
\begin{equation}
    \felas_i = \kappa \Delta \Rij \nij ~,
    \label{eq:felas-M}
\end{equation}
where $\nij$ is a unit vector in the direction of the distance $\bRbij$. Equation (\ref{eq:felas}) is (\ref{eq:felas-M}) expressed in polar coordinates. 
In the overdamped regime, the elastic relaxation time is
\begin{equation}
    \tauelas = \frac{\gamma}{k}~.
\end{equation}
This is the typical decay time of vibrational modes in the absence of hydrodynamic driving force.

The hydroelastic number $\epsilon$  measures the relative significance of elastic and hydrodynamic interactions (\ref{eq:epsilon}),
\begin{equation*}
    \epsilon \equiv \frac{1/\tauelas}{1/\tauhyd} 
    =\frac{\kappa a^3}{\gamma u \, \ell^2}~.
\end{equation*}
Unlike odd viscosity~\cite{Avron1998} or odd elasticity \cite{Scheibner2020}, both hydrodynamic and elastic forces are non-circular, $\nabla_{\Rbij} \cross \fb_{ij} = 0$, as they conserve angular momentum.

\begin{figure*}[htb!]
\centering
\includegraphics[width=1.0\textwidth]
{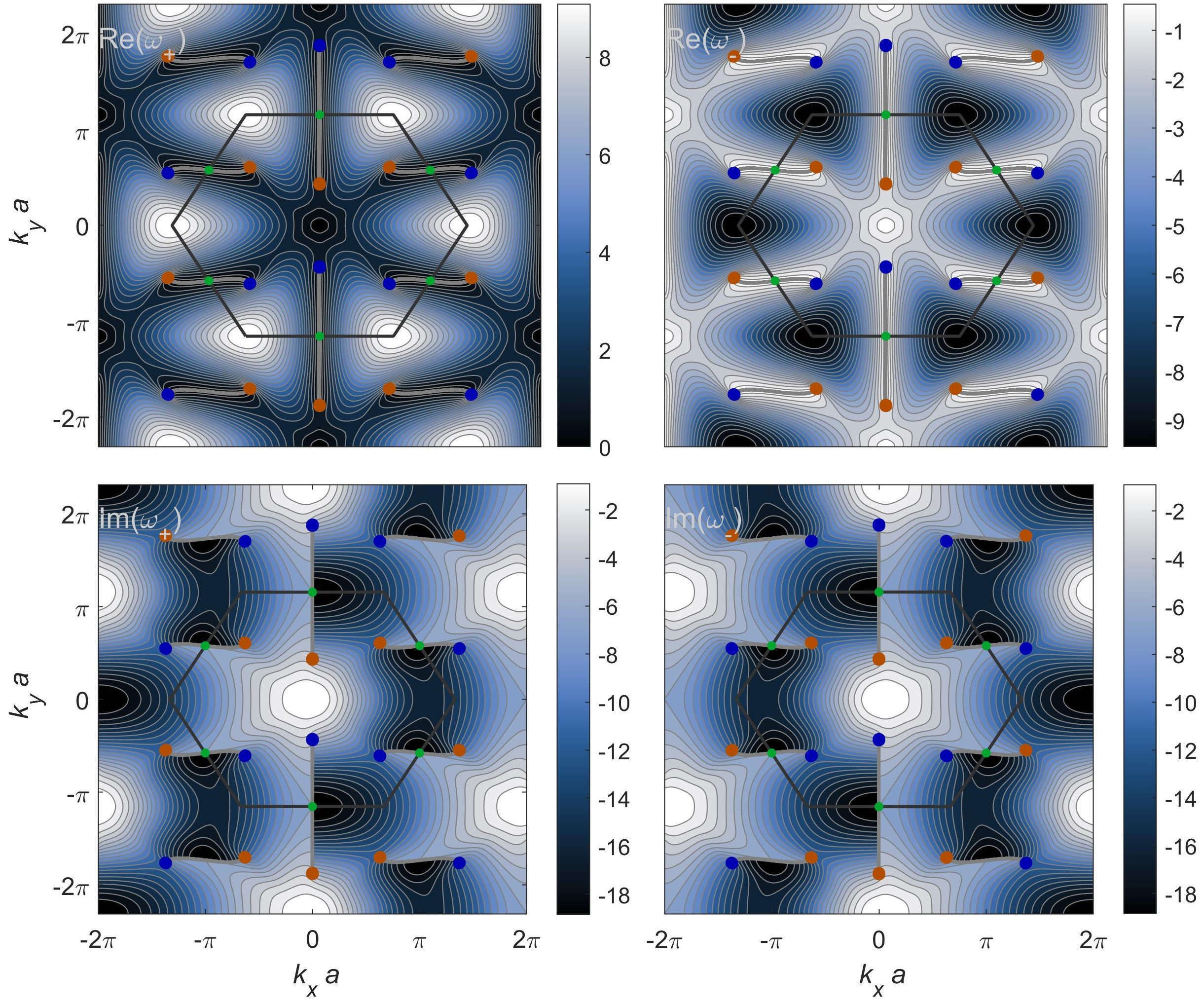}
\caption{\textbf{The spectrum of the triangular lattice.} 
The real (top) and imaginary (bottom) components of the eigenfrequencies $\omega_+$ (left) and $\omega_-$ (right) of a triangular lattice at $\epsilon = \pi$ (Fig.~\ref{fig:2}B).  The black hexagon is the boundary of the Brillouin zone. Shown are the Dirac points (green) at the midpoints of the Brillouin zone edges, and the ExPs with their $+\half$ (red) and $-\half$ (blue) charges. Fermi arcs are grey curves. }
\label{fig:allspec-S}
\end{figure*}

%
\subsection{Linear Expansion}
\label{app:linear_expansion}

In the overdamped low-Reynolds regime, the equations of motion are (\ref{eq:dynamic}). At steady-state, the lattice interactions vanish by symmetry, $\sum_{j \neq i}{(\fhyd_{ij} + \felas_{ij})}=0$, and the lattice moves uniformly at a velocity $u = F/\gamma$. Expansion of the equations of motion (\ref{eq:dynamic}) in small deviations of the lattice positions around the steady-state positions, $\rb_j = \Rb_j - \bar{\Rb}_j$, yields a linear dynamic equation,
\begin{equation}
    \dot{\rb} = \HH \, \rb~,
    \label{eq:linear tensor-M}
\end{equation}
where $\rb$ is $2N$-vector of the $N$ particle deviations, $\rb_i$. The tensor $\HH = \Hhyd + \epsilon \Helas$, which is analogous to a Hamiltonian, combines contributions from hydrodynamics and elasticity. $\HH$ is a $2N \times 2N$-matrix, composed of $2 \times 2$ blocks $\Hij$ that account for interactions between the \ith and \jth particles, 
\begin{equation}
 \Hij  =  \Hijhyd + \epsilon \Hijelas~,   
\end{equation}
where the hydrodynamic term is 
\begin{equation}
    \Hijhyd  = 
    2 \l( \frac{a}{\bRij} \r)^3
    \begin{bmatrix} 
    \cos{3 \tetij} & \sin{3 \tetij}\\ 
    \sin{3 \tetij} & -\cos{3 \tetij}
    \end{bmatrix}~,
    \label{eq:Hijhyd-M}
\end{equation}
and the elastic one 
\begin{equation}
    \Hijelas  =
    \frac{1}{2}
    \begin{bmatrix} 
    1 + \cos{2 \tetij} & \sin{2 \tetij}\\ 
    \sin{2 \tetij} & 1-\cos{2 \tetij}
    \end{bmatrix} ~,
    \label{eq:Hielas-M}
\end{equation}
which are conveniently expressed in terms of Pauli's matrices $\sigx,\sigy,\sigz$ (and $\One$ the unity matrix),
\begin{align}
    \Hijhyd & = 2 \l(a/\bRij\r)^3
    \l( \sin{3 \tetij} \sigx +
    \cos{3 \tetij} \sigz \r)~, 
    \label{eq:Hij-M}\\ 
    \Hijelas &  = \half
    \l( \One + \sin{2 \tetij} \sigx +
    \cos{2 \tetij} \sigz \r)~. \nonumber 
\end{align}
The diagonal terms ensure zero sums, $\HH^{ii} = -\sum_{j \neq i}{\Hij}$.  Since the angles obey $\tetji = \pi + \tetij$,  it follows from (\ref{eq:Hij-M}) that $\Hjihyd$ is odd  with respect to particle exchange ($i {\leftrightarrow} j$),  while $\Hijelas$ is even,
 \begin{equation*}
     \Hjihyd = -\Hijhyd~,~\Hjielas = +\Hijelas~.
 \end{equation*}
The mutual forces between the \ith and the \jth particles are $\fb_{ij} = \Hij (\rb_j -\rb_i)$ and \quad
 $\fb_{ji} = \Hji ( \rb_i - \rb_j)$. Thus, we verify that the hydrodynamic forces violate Newton's third law of momentum conservation, whereas the lastic forces obey it (see Fig.~\ref{fig:1}B),
 \begin{equation*}
    \fhydji  =  +\fhydij ~, \quad
    \felasji =  -\felasij~.
\end{equation*}
While the microscopic molecular forces in the fluid obey Newton's law, the hydrodynamic interactions are \textit{effective macroscopic} forces that do not conserve momentum. The momentum is leaking through the walls and is compensated by the driving force (\eg gravitation or pressure gradient).

\subsection{Momentum space}
\label{app:momentum_space}

To exploit the crystal symmetry, one represents the dynamics in the momentum space of the wave-vectors $\kb = (k_x,k_y)$ by Fourier transform. The linearized dynamical equations (\ref{eq:linear tensor-M},\ref{eq:Hij-M}) are expanded in plane waves, such that the deviation of each particle from its mechanical equilibrium position $\bar{\Rb}_j$ is
\begin{equation}
    \rb_j(t) = \ek e^{i \kb \cdot \bRbj} = 
    \ek e^{i \l( \kb \cdot \bRbj- \omega t \r)}~,
    \label{eq:plane wave-M}
\end{equation}
where $\ek(t) = \ek \exp(-i \omega t)$ is a 2D polarization vector in $\kb$-space. 

\begin{figure*}[htbp!]
\centering
\includegraphics[width=1.0\textwidth]
{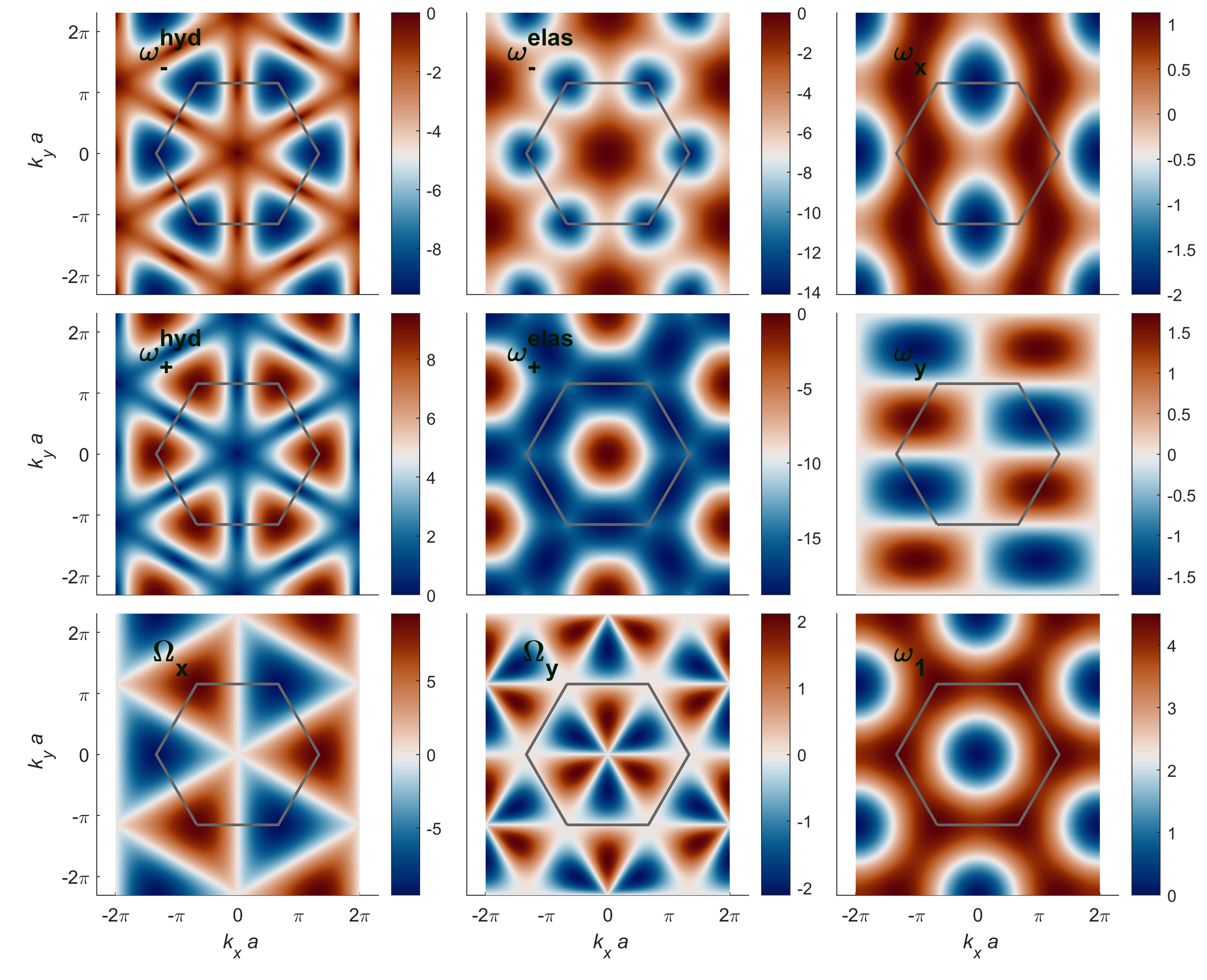}
\caption{\textbf{Hydrodynamic and elastic interactions and frequency bands.} The black hexagon is the boundary of the Brillouin zone. The eigenfrequencies of the purely hydrodynamic systems with negligible elasticity, $\omhyd_\pm$ (\ref{eq:omhyd-M}), are real because the effective Hamiltonian is Hermitian, $\Hk = \Hkhyd$. (top-right and center-right). In the absence of flow, the elastic system is purely damping. $\Hk = \Hkelas$ is skew-Hermitian, and therefore its spectrum, $\omelas_\pm$ (\ref{eq:omelas-M}) is purely imaginary (top-center and center). The elastic interactions, $\Omx$ and $\Omy$ vanish at the Dirac points, the middles of Brillouin zone edges (bottom left and center), while the elastic interactions, $\omx$, $\omy$ and $\omO$, exhibit extrema (left column). 
}
\label{fig:omegas}
\end{figure*}

The equation of motion of $\ek(t)$ is a non-unitary Schr\"odinger-like equation,
\begin{equation}
    i \frac{\partial }{\partial t}\ek(t) = \Hk \ek(t)~,
    \label{eq:schrodinger-M}
\end{equation}
whose eigenvector is $\ek$ with eigenfrequency $\omega$ (\ref{eq:schrodinger}),
\begin{equation*}
        \Hk \ek = \omega \ek~.
\end{equation*}
The ``Hamiltonian" $\Hk$ is a $2\times 2$-matrix expressed in terms of Pauli's matrices as
\begin{align}
    \Hk & = \Hkhyd + \epsilon\, \Hkelas 
    \label{eq:Hk-M}\\ 
    &= \l(\Omx \sigz + \Omy \sigx \r)
    - \epsilon \, i 
    \l(\omO \One +\omx \sigz + \omy \sigx \r)~.
    \nonumber
\end{align}
The contributions of the long-range hydrodynamic to $\Hk$ (\ref{eq:Hk-M}) are Fourier sums,
\begin{align*}
    \Omx = &  -2 i
    \sum_{j \neq 0} 
    \l( \frac{a}{\bRj}\r )^3 \cos{3 \tetj}\,
    e^{ i \kb \cdot \bRbj} ~,  \\
    \Omy = & -2 i 
    \sum_{j \neq 0} 
    \l( \frac{a}{\bRj}\r )^3 \sin{3 \tetj}\,
    e^{ i \kb \cdot \bRbj} ~,
\end{align*}
where $\bRbj = \bR_j(\cos{\tetj},\sin{\tetj})$ are the distances of the steady-state lattice positions from an arbitrary origin particle $0$. Owing to the crystal's parity symmetry, we can rearrange the summation to be over pairs at inverse positions, $\pm \bRbj$, demonstrating that $\Omx$ and $\Omy$ are always real in a crystal,
\begin{equation}
    \begin{bmatrix}
    \Omx \\ \Omy
    \end{bmatrix} =  
    2 \sum_{ j \neq 0 } 
    \l( \frac{a}{\bRj}\r )^3 
    \begin{bmatrix}
    \cos{3 \tetj} \\  \sin{3 \tetj}
    \end{bmatrix}
    \,\sin{\l(\kb \cdot \bRbj\r)} ~.  
    \label{eq:Omegas-M}
\end{equation}
The elastic contributions in (\ref{eq:Hk-M}) are sums over the neighbors connected by springs to particle $0$,
\begin{align*}
    \omx &=  \half \sum_{j \neq 0} \cos{2\tetj}  \,
    \l(1- e^{ i \kb \cdot \bRbj}  \r)~,  \\  
    \omy &= \half \sum_{j \neq 0} \sin{2\tetj}  \,
    \l(1- e^{ i \kb \cdot \bRbj}  \r)~,    \\
    \omO & =  \half \sum_{j \neq 0} 
    \l(1 - e^{ i \kb \cdot \bRbj} \r)~.
\end{align*}
Again, owing to the crystal parity symmetry, the sums can be rearranged, demonstrating that $\omx$, $\omy$ and $\omO$ are all real, 
\begin{equation}
    \begin{bmatrix}
    \omO \\ \omx \\ \omy
    \end{bmatrix} =  
    \sum_{j \neq 0 } 
    \begin{bmatrix}
    1 \\ \cos{2\tetj} \\ \sin{2\tetj} 
    \end{bmatrix}
    \sin^2 \l(\half \kb \cdot \bRb_j\r)~.  
    \label{eq:omegas-M}\\  
\end{equation}
The hydrodynamic and elastic interaction in a triangular lattice are shown in Fig.~\ref{fig:omegas}.

\subsection{Spectra}
\label{app:spectra}

The eigenfrequencies $\omega$ are found by solving the secular equation corresponding to (\ref{eq:schrodinger-M},\ref{eq:Hk-M}).  There are two eigenfrequency bands,
\begin{align}
    \omega_{\pm} = &
    - i \epsilon \,\omO \pm 
    \sqrt{\l(\Omx - i \epsilon \, \omx\r)^2 +
    \l(\Omy - i \epsilon \, \omy\r)^2}~ 
    \nonumber \\
    = & 
    - i \epsilon \,\omO \pm 
    \sqrt{\nu_+ \nu_-}~,
    \label{eq:specturm-M}
\end{align}
where $\nu_{\pm}$ are defined as 
\begin{equation*}
  \nu_{\pm} \equiv
  \l( \Omx - i \epsilon \omx \r) \pm
  i   \l( \Omy - i \epsilon \omy \r)~.
\end{equation*}
In the hydrodynamics-dominated regime, $\epsilon \ll 1$, the spectrum (\ref{eq:specturm-M}) is purely real,
\begin{equation}
 \omhyd_\pm \simeq \, \pm 
 \l( \Omy ^2 + \Omx^2  \r)^{1/2}~,
 \label{eq:omhyd-M}
\end{equation}
while in elasticity-dominated regime $\epsilon \gg 1$ the spectrum is purely imaginary, 
\begin{equation}
    \omelas_{\pm} \simeq \,-i\epsilon
    \l[ \omO\pm (\omx^2+\omy^2)^{1/2} \r]~.
     \label{eq:omelas-M}
\end{equation}
The polarization eignevectors, $\ek^{+}$ and $\ek^{-}$, are 
\begin{align}
    \ek^{\pm} = &
    \begin{bmatrix}
     \Omx - i \epsilon \omx + \omega_{\pm} 
     + i \epsilon \omO  \\
     \Omy - i \epsilon\,\omy
    \end{bmatrix}~    
    \label{eq:eigenvectors-M} \\
    = &
    \begin{bmatrix}
     \Omx - i \epsilon \omx  \pm
     \sqrt{\nu_+ \nu_-}  \\
     \Omy - i \epsilon\,\omy
    \end{bmatrix}~, \nonumber
\end{align}
where the eigenvalues $\omega_{\pm}$ are given in (\ref{eq:specturm-M}). When normalized, the eigenvectors become
\begin{equation}
      \ek^{\pm} = \frac{1}
      {\sqrt{2} \sqrt{ \l| \nu_+ \r| + \l| \nu_- \r|}}
      \begin{bmatrix}
      \sqrt{\nu_{+}} \pm \sqrt{\nu_{-}}\\
      i \l( \sqrt{\nu_{+}} \mp \sqrt{\nu_{-}} \r)
      \end{bmatrix}~.
      \label{eq:normeigenvectors-M}
\end{equation}

\subsection{Circular polarization basis}
\label{app:circular}

One can represent the Hamiltonian in the basis of left and right circularly polarized unit vectors (which are the coalescing eigenstates at the ExPs),
\begin{equation*}
     \frac{1}{\sqrt{2}}
      \begin{bmatrix}
      1\\
      \pm i
      \end{bmatrix}~.
\end{equation*}
In this basis, the Hamiltonian (\ref{eq:Hk-M}) becomes equation (\ref{eq:Hk}),
\begin{align*}
     \Hk & = \Hkhyd + \epsilon\, \Hkelas \\ 
    &= \l(\Omx \sigx + \Omy \sigy \r)
    - \epsilon \,i 
    \l(\omO \One +\omx \sigx + \omy \sigy \r)~.
\end{align*}
The eigenvectors in this representation take the simple form 
\begin{align}
      \ek^{+} = & \frac{1}
      { \sqrt{ \l| \nu_+ \r| + \l| \nu_- \r|} }
      \begin{bmatrix}
      \sqrt{\nu_+}\\
      \sqrt{\nu_-}
      \end{bmatrix}~,~
      \label{eq:CircularEigenvectors-M}\\
      \ek^{-} = &\frac{1}
      { \sqrt{ \l| \nu_+ \r| + \l| \nu_- \r|} }
      \begin{bmatrix}
      \sqrt{\nu_-}\\
      \sqrt{\nu_+}
      \end{bmatrix}~. \nonumber
\end{align}

\begin{figure*}[htb!]
\centering
\includegraphics[width=1.0\textwidth]
{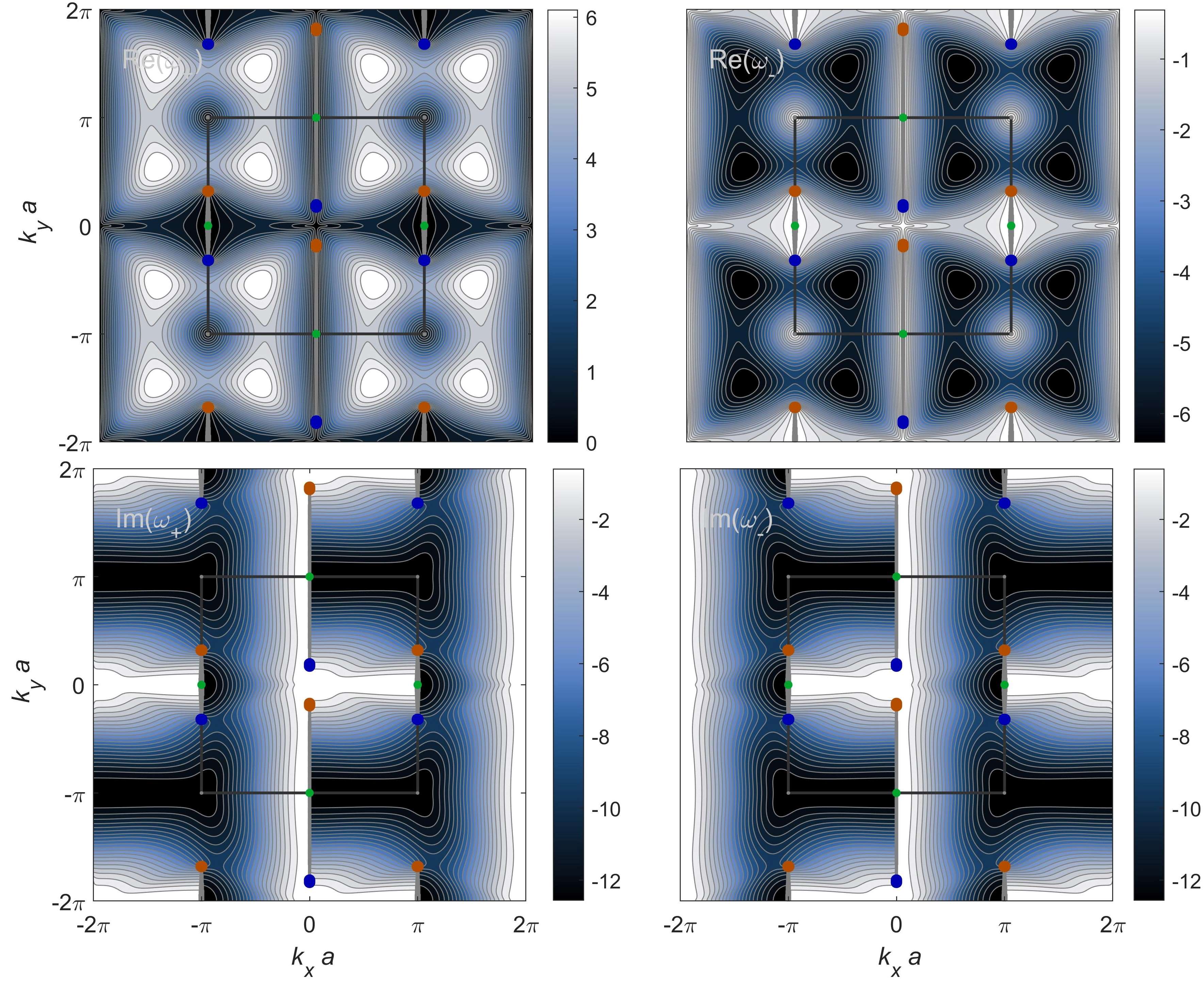}
\caption{\textbf{The spectrum of a square lattice.} 
The real (top) and imaginary (bottom) components of the eigenfrequencies, $\omega_+$ (left) and $\omega_-$ (right), of a square lattice at $\epsilon = \pi$. The black square is the boundary of the Brillouin zone. Shown are the Dirac points (green) and the ExPs with their $+\half$ (red) and $-\half$ (blue) charges. Bulk Fermi arcs are grey curves. }
\label{fig:allS-square}
\end{figure*}

\subsection{Dirac points and cones}
\label{app:Dirac_points}

Dirac points occur in the purely hydrodynamic system ($\epsilon = 0$), at wavevectors $\kb_D$ for which the hydrodynamic interaction vanishes, 
\begin{equation*}
    \Omx = \Omy = 0~.
\end{equation*}
From (\ref{eq:Omegas-M}), one sees that this happens when $\sin(\kb \cdot \bRbj) = 0$, that is for wavevectors that are halves of the reciprocal lattice base vectors, $\half \bb_1, \half \bb_2$, and their combinations, $\kb_D = \half \beta_1 \bb_1 + \half \beta_2 \bb_2$ (where $\beta_1,\beta_2 \in \{-1,0,1 \}$). In the triangular lattice, the six Dirac points are $(\beta_1,\beta_2) = (0,\pm 1), (\pm 1,0), (\pm 1,\pm 1)$. Note that these are the midpoints of the Brillouin zone edges and not the corners as in graphene.

The expansion of (\ref{eq:Omegas-M}) around the Dirac point is linear in $\dkb = \kb - \kb_D$,
\begin{equation*}
    \Omx \sim \nabla_\kb \Omx  \cdot  \dkb ~,~
    \Omy \sim \nabla_\kb \Omy \cdot  \dkb  ~,
\end{equation*}
where the gradients are 
\begin{align}
    \nabla_\kb \Omx  = 
    & \, 2 a \sum_{ j \neq 0 } 
     \l( -1 \r)^{\beta_1 \alpha_1^j + \beta_2 \alpha_2^j }
     \begin{bmatrix}
     \cos{\tetj} \\
     \sin{\tetj}
     \end{bmatrix}
     \frac{\cos{3 \tetj}}{\l(\bRj/a \r)^2}~,
     \label{eq:gradDP-M}\\
     \nabla_\kb \Omy = 
     & \, 2 a \sum_{ j \neq 0 } 
     \l( -1 \r)^{\beta_1 \alpha_1^j + \beta_2 \alpha_2^j }
     \begin{bmatrix}
     \cos{\tetj} \\
     \sin{\tetj}
     \end{bmatrix}
     \frac{\sin{3 \tetj}}{\l(\bRj/a \r)^2}~. \nonumber
\end{align}
The $\alpha_1^j$ and $\alpha_2^j$ in (\ref{eq:gradDP-M}) are the indices of the lattice points, 
$\bRbj = \alpha_1^j \ab_1 + \alpha_2^j \ab_2$, with the basis vectors, $\ab_1$ and $\ab_2$. It is straightforward to verify that the gradients at the Dirac point are orthogonal,
\begin{equation}
    \nabla_\kb \Omx  \cdot
    \nabla_\kb \Omy  = 0~,
    \label{eq:orthGradDP-M}
\end{equation}
resulting in an elliptic cone (Fig.~\ref{fig:2}A). It is also useful to note that, at the Dirac points, the elastic interactions, $\omx$, $\omy$ and $\omO$, 
have extrema (minima, maxima, and saddles), and are therefore constant to first order in $\dkb$ (Fig.~\ref{fig:omegas}).

\subsection{Exceptional points and Fermi arcs}
\label{app:EPs_and_FermiArcs}

Exceptional points (ExPs) occur when both eigenfrequencies (\ref{eq:specturm-M}) and eigenvectors (\ref{eq:normeigenvectors-M}) coincide, 
\begin{align}
    \omega_{+} = \omega_{-} & = - i \epsilon \omO~,\\
    \ek^{+} = \ek^{-}  & = 
    \frac{1}{\sqrt{2}}
    \begin{bmatrix}
    1 \\ \pm i
     \end{bmatrix}~, \nonumber
\end{align}
where the $\pm$ signs correspond to the paired ExPs. In the circular polarization basis, the eignevectors at the ExPs are the unit vectors,
\begin{equation*}
    \ek^{+} =  
    \begin{bmatrix}
    1 \\ 0
     \end{bmatrix}~,~
     \ek^{-} =
    \begin{bmatrix}
    0 \\ 1
     \end{bmatrix}~.
\end{equation*}
ExPs are positioned exactly where the determinant, $\nu_+ \nu_-$, in (\ref{eq:specturm-M}) vanishes. In other words, when $\nu_+ = 0$ or $\nu_- = 0$, which amounts to the double condition (\ref{eq:exceptional}),
\begin{equation*}
    \frac{\Omy}{\omx} = -\frac{\Omx}{\omy} 
    = \pm \epsilon ~.
\end{equation*}
As one varies $\epsilon$, the ExPs move along the 1D curve
\begin{equation}
    \Omx \omx + \Omy \omy = 0~,
    \label{eq:trajectory-M}
\end{equation}
from the Dirac cones at $\epsilon = 0$ to the Brillouin zone corners (or its center) at $\epsilon = \infty$. The stretch of the trajectory between the two ExPs is the \textit{Fermi arc}. Along these trajectories, the eigenfrequencies are
\begin{align}
        \omega_{\pm} & = - i \epsilon \,\omO \pm 
    \l[ \l( \Omx^2 + \Omy^2 \r) - 
    \epsilon^2  \l(\omx^2 +\omy^2 \r) \r]^{1/2} \\
    \label{eq:trajectory-M}
   &  = - i \epsilon \,\omO \pm 
    \l( s^2 - \epsilon^2 \r) ^{1/2}
    \l(\omx^2 +\omy^2 \r)^{1/2}~, \nonumber
\end{align}
where $s \equiv \Omx /\omy = -\Omx/\omy$ is the coordinate along the trajectory, from $\earc = 0$, the Dirac point, through $\earc = \epsilon$, the ExP, to the corner of the Brillouin zone $\earc = \infty$ (the four S-shaped arcs), or its center (the two straight arcs). The exceptional point is the position of a \textit{bifurcation} transition where the branch-cut of the square root stretches along the Fermi arc.

\subsection{Bulk Fermi arcs are generic}
\label{app:generic}

The bulk Fermi arcs observed in the triangular (Figs.~\ref{fig:2},\ref{fig:allspec-S}) or the square (Fig.~\ref{fig:allS-square}) lattices are a \textit{generic} phenomenon of the hydroelastic lattice. They emerge as soon as symmetry is broken, at $\epsilon \neq 0$, when the skew-Hermitian elasticity is introduced. 
To see this formally, one may examine the emergence of arcs for small $\epsilon$.
Then, the exceptional point condition (\ref{eq:exceptional}) can be linearly expended in small deviations from the Dirac point, $\dkb = \kb - \kb_D$  (\ref{eq:gradDP-M}). The double condition (\ref{eq:exceptional}) becomes a linear equation in $\dkb$,
\begin{equation}
\begin{bmatrix}
    \nabla_\kb \Omx ^\intercal \\
    \nabla_\kb \Omy ^\intercal
\end{bmatrix}
\dkb
= \pm \epsilon \,
\begin{bmatrix*}[r]
    - \omy \\ \omx
\end{bmatrix*}~,
\label{eq:smallArc-S}
\end{equation}
where the $\pm$ signs correspond to the two ExPs. Since the gradients are linearly-independent (\ref{eq:orthGradDP-M}), solutions always exist, and the positions of the ExPs are
\begin{equation}
\dkb
= \pm \epsilon\,
\begin{bmatrix}
    \displaystyle
    \frac{\nabla_\kb \Omx}
    {\lvert \nabla_\kb \Omx \rvert^2} \quad 
        \frac{\nabla_\kb \Omy}
    {\lvert \nabla_\kb \Omy \rvert^2} 
\end{bmatrix}
\begin{bmatrix*}[r]
  - \omy \\  \omx
\end{bmatrix*}~.
\label{eq:smallArcSol-M}
\end{equation}
Hence, the arcs and the ExPs are generic. The bulk Fermi arc stretches between the two solutions, and its size grows linearly with $\epsilon$.

\subsection{Berry's phase and the topological charges}
\label{app:Berry}

In the momentum representation, Berry's connections are the vectors~\cite{Berry1984},
\begin{equation}
    \AA_\pm(\kb) = i \,\ek^{\pm\ast} \nabla_\kb \ek^{\pm}~,
    \label{eq:Berry-M}
\end{equation}
corresponding to the eigenvectors $\ek^{\pm}$. For convenience, we write the normalized eigenvectors as Jones polarization vectors,
\begin{equation}
    \ek^{+} = \begin{bmatrix}
    \sqrt{c} \, e^{i \alpha_+}\\
    \sqrt{1-c} \, e^{i \alpha_-}
    \end{bmatrix} ~,~
    \ek^{-} = \begin{bmatrix}
    \sqrt{1-c} \, e^{i \alpha_-}\\
    \sqrt{c} \, e^{i \alpha_+}
    \end{bmatrix} ~.
\end{equation}
In this notation, we find that the Berry connections are simply
\begin{equation}
        \AA_{\pm}(\kb)= 
    - c \nabla_\kb \alpha_+ - (1-c) \nabla_\kb \alpha_- ~.
    \label{eq:BerryJones-M}
\end{equation}
For the eignevectors (\ref{eq:eigenvectors-M}), the Jones parameters are
\begin{align*}
    c  = &
    \frac{\l| \sqrt{\nu_{+}} + \sqrt{\nu_{-}} \r|^2} 
      {2 \l( \l| \nu_+ \r| + \l| \nu_- \r| \r)}~,~ \\
    \alpha_+ = & \arg \l( \sqrt{\nu_{+}} + \sqrt{\nu_{-}} \r)~, \\
    \alpha_- = & \arg \l( \sqrt{\nu_{+}} - \sqrt{\nu_{-}} \r)~.
\end{align*}
The Berry phases $\gamma_\pm$ are the path integrals
\begin{equation}
    \gamma_\pm = \oint \AA_{\pm}(\kb) \cdot \dkb ~.
    \label{eq:BerryPhase-S}
\end{equation}
To find the Berry phase of the ExPs, we integrate over a small circular counterclockwise path around each point. For example, consider the point where $\nu_+ = 0$. In the vicinity of an ExP, $c \simeq \half$, and the connections are therefore,
\begin{equation}
        \AA_{\pm}(\kb) \simeq
        - \half\nabla_\kb \alpha_+  
        - \half \nabla_\kb \alpha_- \simeq
        - \half \nabla_\kb \arg{\nu_+} ~.
\end{equation}
 Along the circular integration path, the phase $\arg{\nu_+}$  passes all quadrants. In particular, it passes  through the branch cut, $\arg{\nu_+} = \pm \pi$, 
 where it jumps by $ \pm 2 \pi$,  depending on the gradients of $\nu_+$.  Thus, we find that Berry's phase is $\gamma_\pm =  \pm \pi$.  Likewise, at the other ExP, Berry's phase  takes the opposite sign,  $\gamma_\pm = \mp \pi$. Altogether, we find that the corresponding topological charges are  $q_\pm = \gamma_{\pm}/(2 \pi) = \pm \half$. \\

\subsection{Vorticity and Topological Charges}
\label{app:vorticity}
 Another way to find the topological charges is through  the vorticity of the eigenfrequency band-gap, $\Delta \omega = \omega_+ - \omega_-$, defined as~\cite{Shen2018}
 \begin{equation}
     \mathcal{V} = -\frac{1}{2\pi} \oint_\Gamma 
     \dkb \cdot \nabla_\kb \arg{\l(\Delta \omega\r)}~,
     \label{eq:vorticity-M}
 \end{equation}
 where $\omega_\pm$ are the given in (\ref{eq:Omegas-M}).  Since, $\Delta \omega = \omega_+ - \omega_- = 2 \sqrt{\nu_+ \nu_-}$, we find that the vorticity is
 \begin{equation}
     \mathcal{V} = -\frac{1}{4\pi} \oint_\Gamma 
     \dkb \cdot \nabla_\kb \l( \arg{\nu_+} + \arg{\nu_-} \r)~.
     \label{eq:vorticityCirc-S}
 \end{equation}
 As in the case of the Berry phase, the vorticity around the ExPs  is determined by their branch-cuts. Namely,  the jumps of $\arg \nu_\pm$ by $\pm 2 \pi$ when passing through the branch cuts yield the topological charges $q_\pm = \pm \half$.

\begin{figure*}[htbp!]
\centering
\includegraphics[width=1.00\textwidth]
{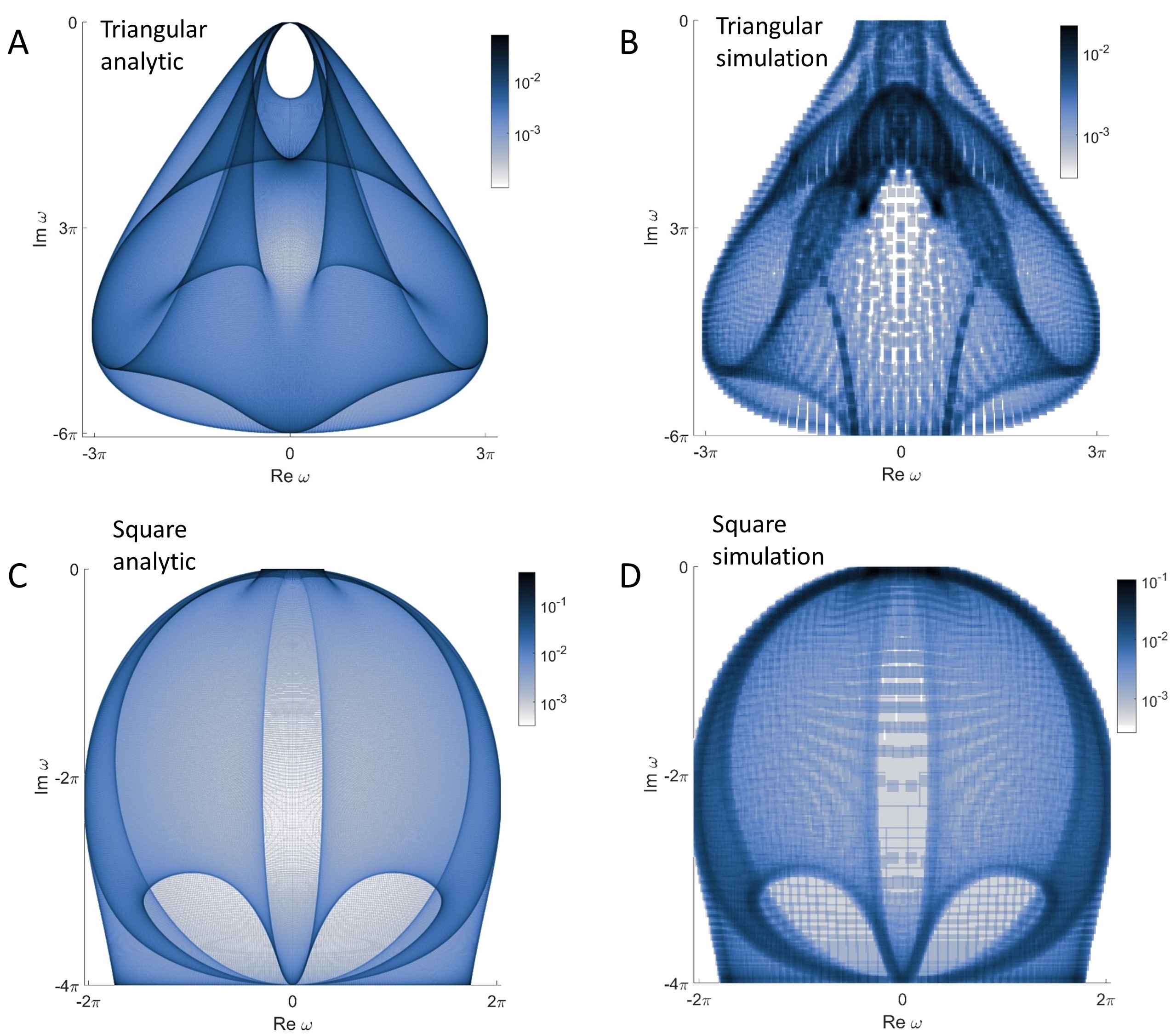}
\caption{
\textbf{Density of states $g(\omega)$: analytic solution vs. simulation.}
(\textbf{A-B}): $g(\omega)$ for triangular lattice, analytic (A) and simulation (B), for $\epsilon = \pi$. The simulated system is a $121 \times 121$ triangular lattice with periodic boundary conditions (a torus). The spectrum is obtained by Fourier analysis of the motion computed from the dynamical equation (\ref{eq:dynamic}). The simulated $g(\omega)$ exhibits general similarity to the analytic solution. The incommensurately of the four-fold symmetry of the torus and the six-fold symmetry of the lattice results 
(\textbf{C-D}): $g(\omega)$ for square lattice, analytic (A) and simulation of a $121 \times 121$ lattice (B), for $\epsilon = \pi$. The correspondence between the simulation and the analytic solution is excellent, owing to the commensurable four-fold symmetry of the lattice and the periodic boundary conditions, with some coarseness due to the finite size of the simulated system.}
\label{fig:simulations-S}
\end{figure*}

\clearpage
\bibliography{NonHermitian}

\end{document}